\documentclass[11pt]{article}

\makeatletter

\newcommand{\Rmnum}[1]{\expandafter\@slowromancap\romannumeral #1@}
\makeatother
\makeatletter
\normalsize
\usepackage{epsfig}
\usepackage{url}
\usepackage{amsmath}
\usepackage{chemarr}
\usepackage{fixltx2e}
\usepackage{amsthm}
\usepackage[version=3]{mhchem}
\usepackage{amsfonts}
\usepackage{indentfirst}
\usepackage{graphics,color}
\usepackage{acronym}
\usepackage{txfonts}
\usepackage{floatflt}
\usepackage{setspace} 

\graphicspath{{./Figures/}}

\usepackage{lipsum}
\usepackage{floatrow}

\newcommand{\comment}[1]{}

\acrodef{rbs}[RBS]{ribosome binding site}
\acrodef{rnaps}[RNAPs]{RNA polymerases}
\acrodef{rnap}[RNAP]{RNA polymerase}
\acrodef{gfp}[GFP]{green fluorescent protein}
\acrodef{sal}[SAL]{salicylate}
\acrodef{ahl}[AHL]{\textit{N}-hexanoyl-{\scriptsize\textsc{L}}-homoserine lactone}
\acrodef{iffl}[IFFL]{incoherent feed-forward loop}
\acrodef{tf}[TF]{transcription factor}
\acrodef{rfp}[RFP]{red fluorescent protein}
\acrodef{ecf}[ECF]{extracytoplasmic function}
\acrodef{tir}[TIR]{translation initiation region}
\acrodef{aris}[ARIS]{admissible reference input set}
\acrodef{utr}[UTR]{untranslated region}

\begin{document}
\begin{center}
 {\bf\Huge Control systems for synthetic biology and a case-study in cell fate reprogramming}\\

 {\large {D}omitilla Del Vecchio}\\
 {\large Department of Mechanical  Engineering}\\
  {\large Department of Biological  Engineering}\\
  {\large Massachusetts Institute of Technology}\\
 January 27, 2026
\end{center}

%\dois{}{}

{\bf Summary.} This paper gives an overview of the use of control systems engineering in synthetic biology, motivated by  applications such as  cell therapy and cell fate reprogramming for regenerative medicine. A ubiquitous problem in  these and other applications is the ability to control the concentration of specific regulatory factors in the cell accurately despite environmental uncertainty and perturbations.   The paper  describes the origin of these perturbations and how they affect the dynamics of the biomolecular ``plant'' to be controlled. A variety of biomolecular control implementations are then  introduced to achieve robustness of the plant's output to perturbations  and are grouped into feedback and feedforward control architectures.  Although sophisticated control laws can be implemented in a computer today, they cannot be necessarily implemented inside the cell  via biomolecular processes.%, just like initially controllers were implemented with capacitors, resistors, and inductors. 
This fact constraints the set of feasible control laws  to those realizable through  biomolecular processes that  can be engineered   with synthetic biology.  After reviewing biomolecular feedback and feedforward control implementations, mostly focusing on the author's own work, the paper illustrates the application of such control strategies to cell fate reprogramming. Within this context, a master regulatory factor  needs to be controlled at a specific level inside the cell in order to reprogram skin cells to pluripotent stem cells. The article closes by highlighting on-going challenges and directions of future research for biomolecular control design.

%Please refer the IEEEtran\_HOWTO.pdf is the complete guide of \LaTeX\ for manuscript preparation included with this stuff.

\section{Introduction}
\begin{floatingfigure}[r]{9.5cm}
{\hspace{0cm}\includegraphics[width=0.85\textwidth]{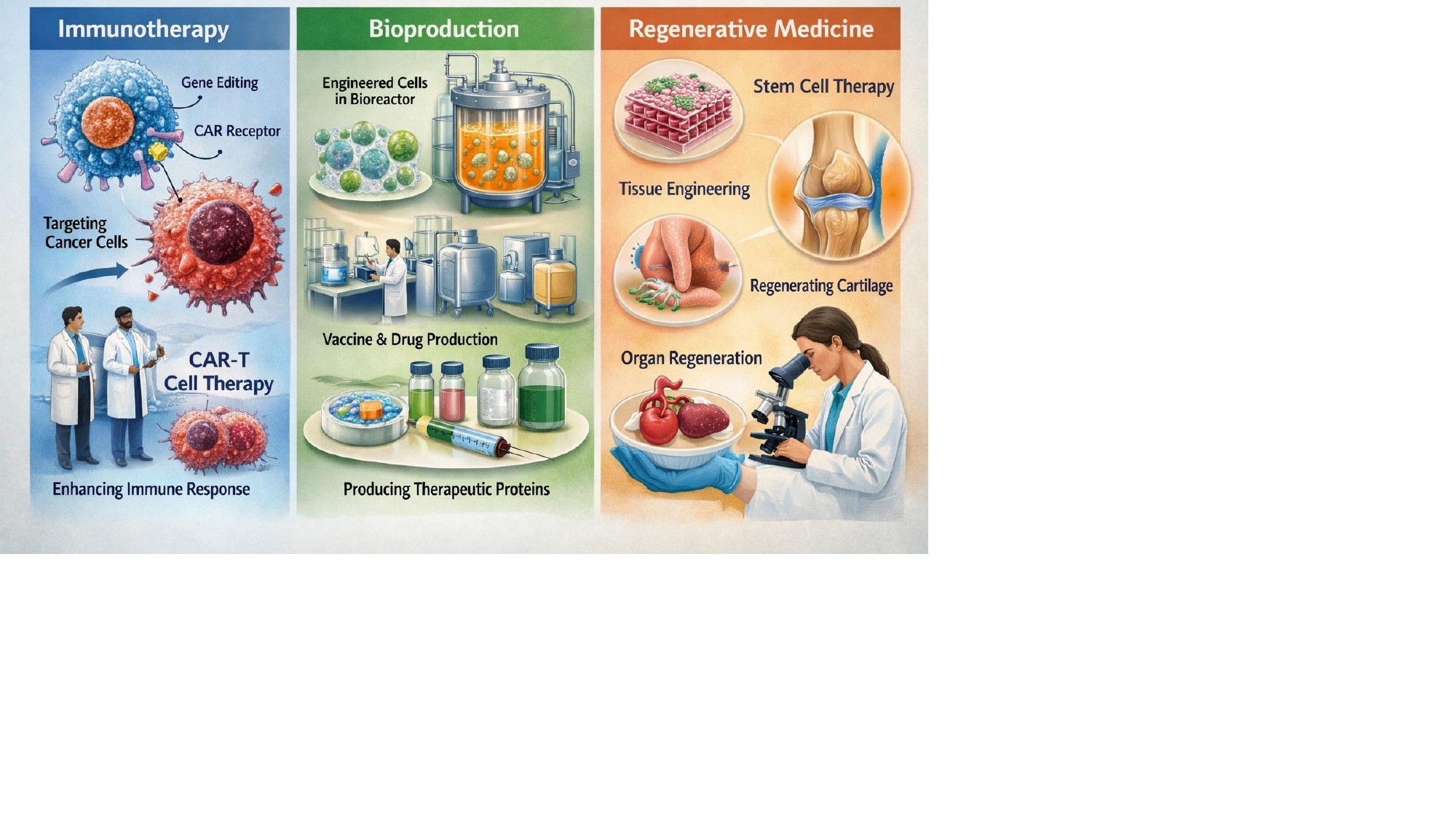}}
\vspace{-3.5  cm}
\caption{ Example Applications of engineered mammalian cells. Left: patient-specific  CAR-T cells can be engineered and generated starting from a patient's own cells for immunotherapy \cite{TCell}. Mammalian cells can be engineered to produce high-value compounds such as monoclonal antibodies \cite{bioproduction}. A patient's damaged tissues can be regenerated using patient-specific induced pluripotent stem cells (iPSCs) \cite{celltherapy}.
\label{fig:0}}
\end{floatingfigure}

Synthetic biology was born in year 2000 when the first two synthetic genetic circuits were built in bacteria: a negative feedback oscillator termed the {\it repressilator} \cite{repressil} and a bistable positive feedback network called the {\it toggle switch} \cite{toggle}. The field has started initially in prokaryotic cells but is  making significant contributions both in the context of bacterial and   mammalian applications \cite{Khalil:2010aa,CameronBashorCollins14}. In particular, mammalian synthetic biology has the potential to impact applications where sensing, computation, and actuation are required, such as in the engineering of T cells for immunotherapy \cite{TCellSynBio2017}, and in applications where  it is important to optimize or fine tune the levels of specific molecules in the cell, such as for bioproduction \cite{ProductionSynBio2022}  and cell fate reprogramming for regenerative medicine \cite{ReproSynBio2020} (Fig. \ref{fig:0}).  Although control and dynamical systems theory have been pivotal for designing the qualitative dynamics of synthetic gene networks since the inception of synthetic biology \cite{hsiao2018control,JRSI2016},    only more recently   control systems approaches have been employed to engineer biomolecular controllers that achieve robust regulation of an output of interest \cite{huang2018quasi,ddv-NBT,Jones2020,Jones2022,dCas9Regulator,KhammashIntegral,frei2021adaptive,BarajasReview,KhammashRes}.  The need for robust output regulation comes from the requirement to achieve specific levels of regulatory molecules inside the cell against perturbations coming from poorly known and variable intra-cellular and extra-cellular environment.   

%(*) Overview of synthetic biology and its main applications and the impact of control engineering in the field – histogram chart – many issues are lack of robustness of genetic circuits to pertrubations, whether they come from inside or outside the cell. 

%
\begin{floatingfigure}[r]{8.5cm}
{\hspace{-1cm}\includegraphics[width=1.3\textwidth]{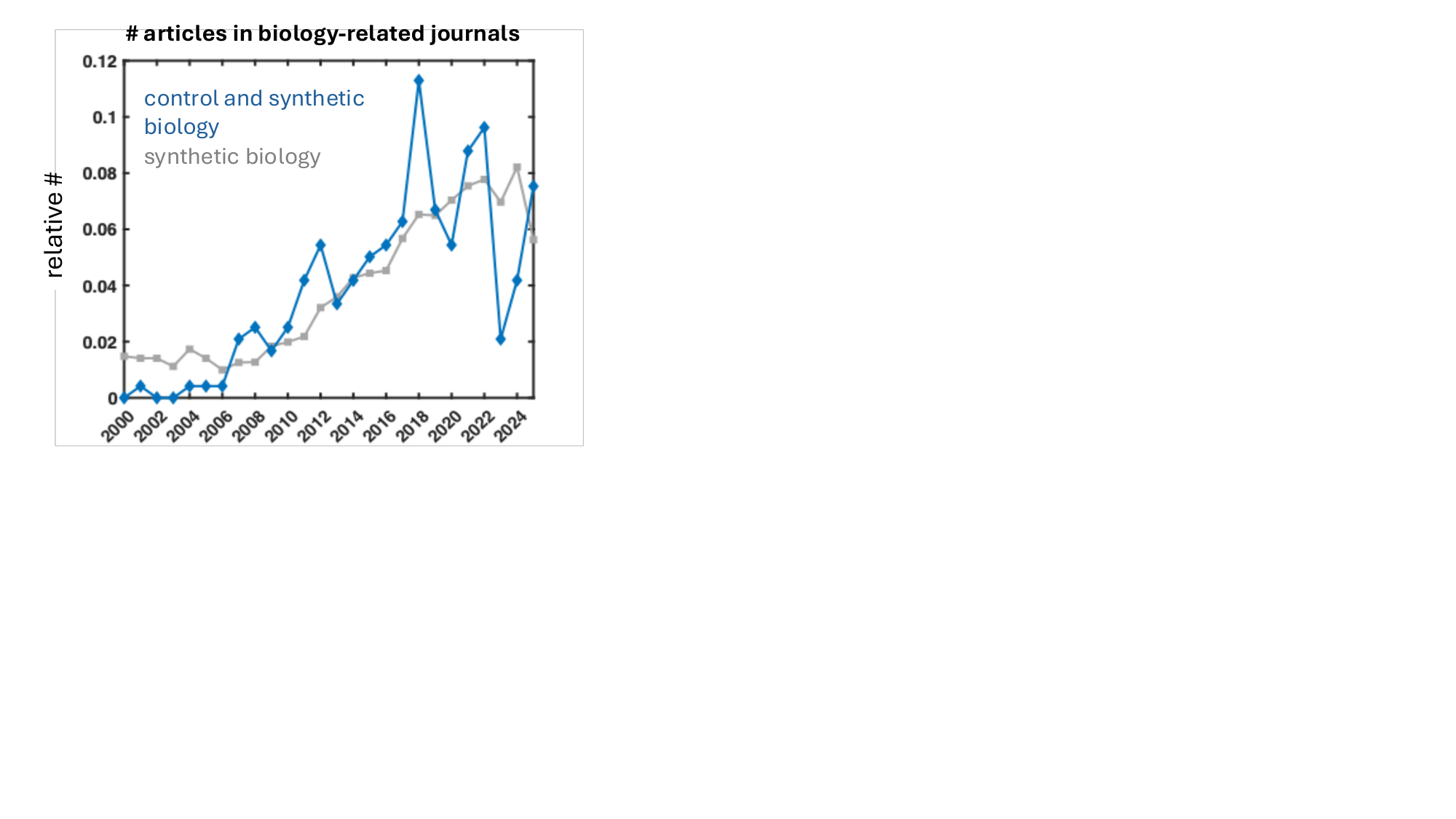}}
\vspace{-6cm}
\caption{ Number of  articles published every year in a biology or bioengineering journal. Numbers are normalized by the total number of articles over the considered time span. The gray line shows synthetic biology articles while the blue line shows articles with both synthetic biology and control systems content. %(b) A genetic device constitutes the simplest input/output module in gene regulatory networks (GRNs), which is constituted by the process of gene expression where a gene produces an output protein (or factor) through the process of transcription and translation \cite{ref}. A synthetic genetic device's output is influenced by several factors, including changes to the cellular environment (context),  to the extracellular context, and directly or indirectly by other genetic devices. The figure has been adapted from \cite{Shakiba2021}.
\label{fig:1}}
\end{floatingfigure}
%
%need to achieve specific levels
Regulatory molecules such as transcription factors (TFs) cover critical roles inside the cell, such as directing the fate of the cell from the pluripotent undifferentiated state to highly specialized cell types such as skin cells, blood cells, and brain cells. Because of their role, steering their concentration to levels that direct cell fate changes has been a cornerstone of cellular reprogramming \cite{ForcingcellstoChangelineage}. For example, by artificially increasing the levels of pluripotency transcription factors  in somatic cells makes it possible to revert the cells to induced pluripotent stem cells (iPSCs) \cite{YamanakaReview}, a discovery that was awarded a Nobel Prize \cite{yama2006}. However, it is also known that the efficiency of this process and the quality of iPSCs generated is linked to the  cellular levels of these TFs \cite{papa}, thereby igniting the question of how it is possible to accurately control the levels of TFs inside the cell instead of injecting them indiscriminately.   A similar situation is found when artificially converting iPSCs to hematopoietic stem cells (HSCs), from which all blood cell types can be derived, thereby providing a regenerating bank of blood cells that can be used to treat a variety of diseases \cite{HSCTherapy}. A path to this conversion involves the differentiation of  iPSCs to the hemogenic endothelium state, a step that   requires specific dosage and timing of lineage-related TFs \cite{LeeD2011}. 
 
In the context of  cell therapy, cells' chromosome is engineered to artificially produce specific regulatory factors that enhance specific cellular functions or enable the cell to produce drugs. These cells  are then delivered to the patient as therapeutic agents \cite{celltherapy}. One example is the case of  patient-specific CAR-T cells for immunotherapy \cite{TCell}, which are engineered starting from a patient's own T cells  or by generating them form the patient's own derived  iPSCs. %Generation of T cells from iPSCs is particularly promising because iPSCs continuously replicate, so they provide a regenerating bank of immune cells.
The engineering of CAR-T cells involves the artificial expression of chimeric antigen receptors (CARs) and the modulation of its level to optimal concentrations for best performance \cite{CARLevel}. Similarly, cell therapy for Parkinson's disease    involves delivering to the patient engineered cells expressing the glial cell line-derived neurotrophic factor (GDNF), a natural protein that supports the survival and growth of neurons, particularly the dopamine-producing cells lost in Parkinson's \cite{GDNF-Parkinson}. This factor, however, needs to be kept at tightly controlled concentrations, which are also patient-specific and likely time-varying \cite{GDNF-Level}.

%(*) General problem of controlling endogenous TF levels as a way to steer cell fate:
%- hiSCs reprogramming: OCT4 intermediate level – from hiPSC paper
%- differentiation into HSCs (SLGE) or endothelial cells ETV2 intermediate level – take it from NIH grant
%- T cell derivation patient specific: balance of RunX3 and ThPOK to achieve balance of helper versus killer cells: In the context of differentiating human induced pluripotent stem cells (hiPSCs) into T helper cells, maintaining a precisely regulated, physiological level of ThPOK expression is crucial for functional outcomes. Expression that is "too high" would likely disrupt the normal, regulated process, potentially leading to non-functional or malignant T cells, thereby hampering the desired differentiation into healthy T helper cells
%- cell therapy: Parkinson – need to have a certain level of the factor GDNF: cells hiPSCs are differentiated to neural cells and used as targeted drug delivery agents. The level of GDNF is important for this – too high leads to side effects, so a middle-low constant level is necessary

%sources of uncertainty
In all these applications, the levels of regulatory molecules should be steered towards optimal levels, but they are affected by the cellular environment in many incompletely known ways. First of all, these molecules are part of a gene regulatory network (GRN) and therefore their expression is regulated by other factors, some of which are known but many others are not known. In addition to this, these molecules are generated by the process of gene expression, which goes through transcription, translation, and post-translational modifications. Each of these steps requires cellular resources, including molecular complexes and enzymes required for transcription, translation, and proper protein folding.  The level of these global, yet limited, resources is affected by the state of the cell, which changes during cell fate transitions, and is also affected by the introduction of artificial genes that express the TFs of interest, since production of these will sequester  the resources \cite{Jones2020,gyorgy2015isocost,Ted-Modular}. All these perturbations affect the level of the TFs that need to be controlled at specific concentrations, thereby the need for robust control arises \cite{Shakiba2021}. Indeed, there has been a growing body of research where control design tools are applied or adapted within the field of synthetic biology as it can be appreciated by the increasing number of publications that use control design tools in biology or bio-engineering journals over the years (Fig. \ref{fig:1}).

%COMMUNITY PICTURE?

\section{The biomolecular ``plant'' and its environment}
\begin{figure*}[t!]
\includegraphics[width=1.1\textwidth]{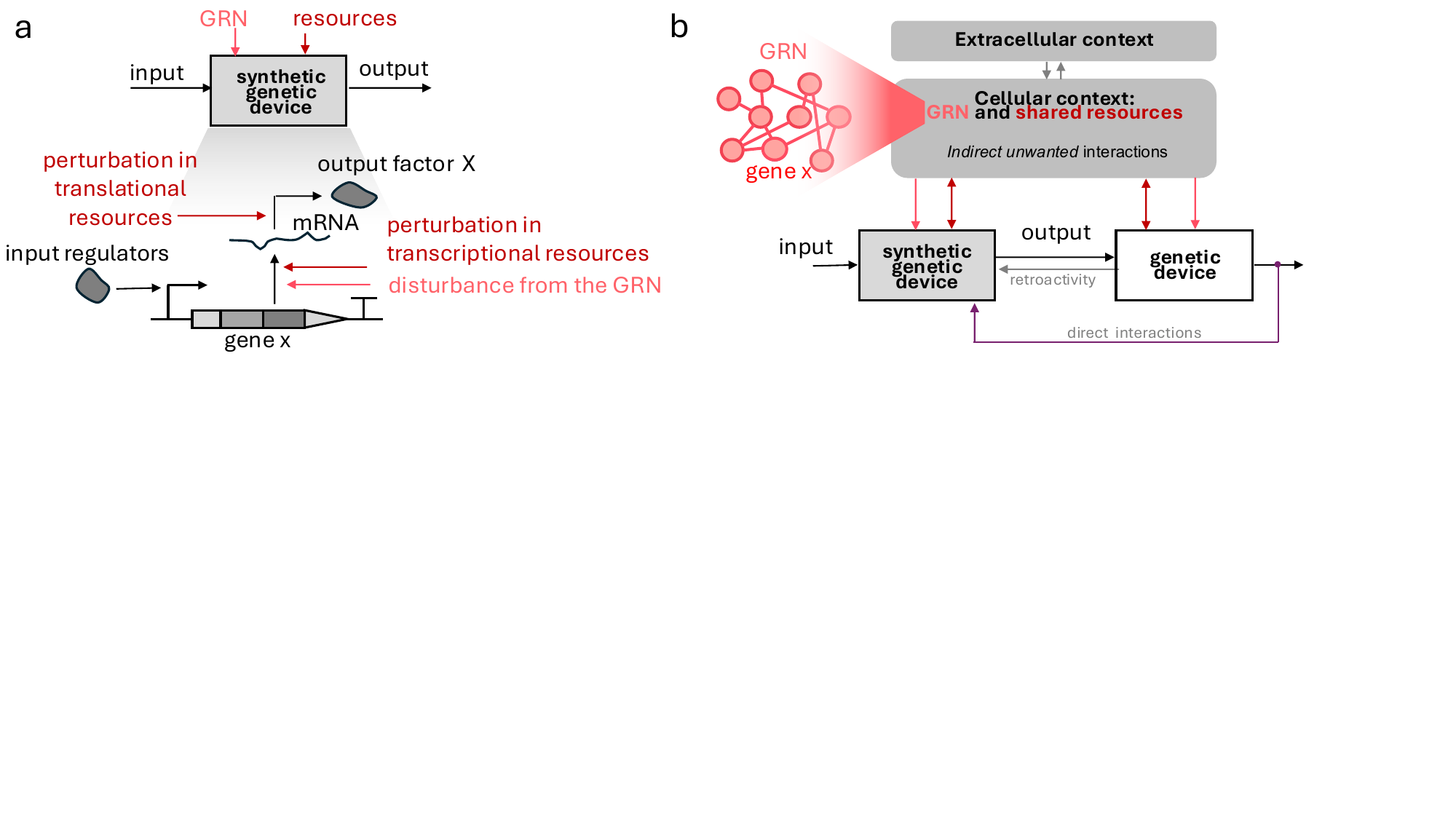}
\vspace{-6cm}
\caption{ Synthetic genetic device and its environment. (a)  A synthetic genetic device is an input/output module constituted by the process of gene expression where a gene produces an output protein  X through the process of transcription and translation \cite{BFS}. When X is a copy of an endogenous protein, the overall production of mRNA is influenced by the endogenous GRN including gene x.  Transcription and translation are affected by perturbations to cellular resources. (b)  Perturbations to cellular resources arise due to changes in the cellular or extra-cellular context and to the sequestration of these resources by other genetic devices  \cite{TCONQian}. The figure has been adapted from \cite{Shakiba2021}.
\label{fig:2}}
\end{figure*}

When the cellular level of an endogenous protein X needs to be artificially controlled, a common approach is to introduce a synthetic genetic device producing protein X in the cell. The genetic device is an input/output module  that is encoded in the DNA through the  gene x that expresses (i.e., produces)  protein X (Fig. \ref{fig:2}a). Gene expression occurs through two steps  according to the central dogma of molecular biology: transcription and translation \cite{BFS}. In the first step, the gene is transcribed into a messenger RNA (mRNA) by means of cellular transcriptional resources, such as the RNA polymerase, transcriptional co-factors and co-activators, and elongation factors. In the second step, the mRNA is translated into an amino-acid chain, which then is folded into a protein, by means of cellular translational resources, such as the ribosome.   

Because protein X is a copy of a  protein that is  also produced by an endogenous gene as part of a gene regulatory network (GRN),  the overall rate of transcription of the mRNA of protein X has both the contribution of the synthetic genetic device, on which we have control, and of the endogenous gene expression, over which we do not have control. We therefore can regard the contribution of the endogenous GRN to the transcription rate as a disturbance. Similarly, the level of cellular resources required for transcription and translation depend on the state of the cell and their availability to the synthetic genetic module is affected by the presence and activation of other genetic modules in the cell, whether synthetic or endogenous \cite{gyorgy2015isocost,Jones2020}.  As a consequence, we can regard the level of these resources also as being uncertain and subject to perturbations (Fig. \ref{fig:2}ab). 

Finally, when the output of the synthetic genetic device regulates its targets, whether in a synthetic or endogenous genetic device, retroactivity arises, which sequesters the regulators from other processes, with often counter-intuitive outcomes \cite{Gyorgy:2014aa}. This can also be regarded as a disturbance on the genetic device, whose effects have been extensively characterized in the literature and mitigation strategies have been developed and reviewed before \cite{BFS}. We therefore focus our attention on the problem of mitigating the effect of perturbations due to the GRN and to the variability of cellular resources in order to achieve a desired level of the factor X in the cell.

To this end, it is useful to introduce a simple mathematical model for the dynamics of the concentration of the mRNA of factor X and of factor X itself, which will represent the ``plant'' to be controlled. Specifically, we let $m$ denote the concentration of the mRNA of factor X and we let $X$ represent the concentration of factor X. With this notation, we can write the rate of change of these concentrations as driven by production and decay processes due to degradation or dilution as follows \cite{BFS}:
\begin{equation}
\frac{dm}{dt}= k-\delta\cdot m+{\color{red}H_{GRN}},\;\;\; \frac{dX}{dt}= \kappa\cdot m- \gamma\cdot X + {\color{red}r},\;\;\; y=X
\label{eq:plant}
\end{equation}
in which $\delta$ and $\gamma$ are positive decay rate constants, $\kappa$ is a positive translation rate constant,  $k$ is a positive transcription rate constant,    $H_{GRN}$ represents the disturbance to the transcription rate due to endogenous transcription of X, and $r$ represents retroactivity arising from X reacting with its targets \cite{Gyorgy:2014aa}. The production rate constants $k$ and $\kappa$ are proportional to the level of free cellular resources required for transcription and translation and take the following general form \cite{BFS}:
\begin{equation}
k=\alpha \cdot {\color{red}R_{TX}},\;\;\;\kappa=\beta \cdot {\color{red}  R_{TL}},
\label{eq:plant1}
\end{equation}
in which  $\alpha>0$ can  be a function of the concentration of transcriptional regulators u, if present, and takes the form of  a Hill function \cite{BFS}, $\beta>0$, $R_{TX}$ represents the concentration of free transcriptional resources, such as RNA polymerase, and $R_{TL}$ represents the concentration of free translational resources such as the ribosome. These free resource levels are subject to a change when the demand for them by other genetic modules changes \cite{gyorgy2015isocost,qian2017resource,Jones2020} or the cellular state or environment changes \cite{Jones2020}. Therefore, they will be considered as constant parameters except for cases when other modules are being activated or when the cellular state changes. % disturbance on transcriptional co-factors resulting from variable demand for them by other genetic devices (endogenous or synthetic), and similarly $d_{TL}>0$ represents a disturbance on translational resources resulting from demand from them from synthetically or endogenously produced mRNAs. The forms in which these perturbations enter the genetic device is obtained by writing the conservation laws for resources in each of the transcription and translation steps and by solving for the amount of free resources, transcriptional co-factors or ribosomes. For more details on the form of $k$ and its experimental validation, the reader is referred to the supplementary file of \cite{Jones2020}, which details models for transcription co-factor competition in mammalian cells. A simplified version of these models is also available as an on-line supplement to \cite{BFS}. For details on the derivation of the form of $\kappa$ and its experimental validation, the reader is referred to \cite{Isocost,ACSSynBio2017,QianTCON}. 

In addition to resources required for gene expression, also the processes of mRNA degradation and protein degradation rely on cellular resources, such as RNAses for mRNA and proteases for proteins \cite{BFS}. As a consequence, the decay rate constants $\delta$ and $\gamma$ may also be subject to change if the demand for these degradation resources changes. While the extent of perturbation in gene expression resources was experimentally shown to be major and is currently a main hurdle in predictive design of genetic circuits \cite{Ted-Modular}, it is less clear the extent  to which  perturbations to degradation resources is significant and is still subject of investigation. Therefore, it is not considered in this model. Finally, parameter $\alpha$ encodes the number of DNA copies of the genetic module, which is one copy if the genetic module is encoded in the genome via site-specific chromosomal integration, but it is multiple copies when the genetic module DNA is delivered via transient transfection or via viral integration. In this latter case, this parameter is also variable across cells.  In summary, for this paper, the system of ordinary differential equations (ODEs) (\ref{eq:plant}) with expressions (\ref{eq:plant1}) represents a simple model of the ``plant'' with output of interest $y=X$, with the effect of the environment captured by  $H_{GRN}, r$, and possible variability in $R_{TR}$,   $R_{TL}$, and $\alpha$.

This simple mathematical model, being deterministic, assumes that molecular counts in the cell are sufficiently  high such that it is possible to define concentrations  as counts divided by the cell volume \cite{BFS}. Also, this model assumes that the cell volume is ``well mixed'', thereby neglecting the fact that mammalian cells have a nucleous and that the mRNA migrates from the nucleus to the cytoplasm for being translated, through a process called splicing \cite{Albert}. Nevertheless, this simple model has been widely used and shows satisfactory agreement with experimental data. For stochastic models of gene expression, the reader is referred to other references \cite{Swain:2002ww}. In this paper, the above model will be   used as a guidance to illustrate the development of biomolecular controller design for disturbance attenuation.

%(*) GRN interference

%(*) high-level context dependence

%(*) resource changes and cell state changes general cellular TX/TL co-factors, like mediator or ribosomes.

%(*) retroactivity and off-target effects

%(*) closeup view of process with control inputs and external distrubances

\section{Biomolecular controller design for disturbance attenuation}\label{sec:integral}
\begin{figure*}[t!] 
\includegraphics[width=1\textwidth]{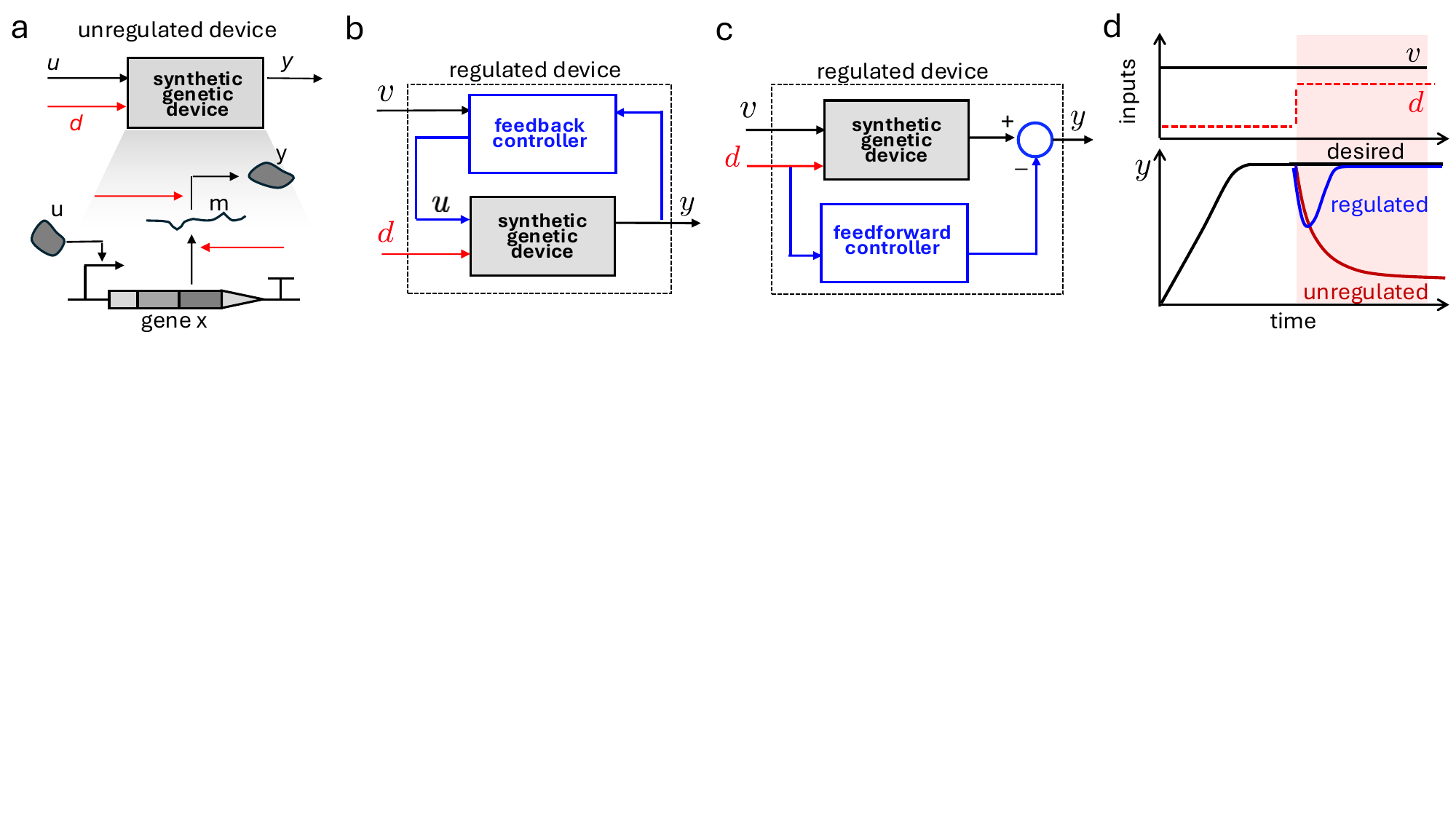}
\vspace{-6cm}
\caption{Biomolecular controller architectures for a synthetic genetic device. (a) Unregulated genetic device with disturbances acting either on transcription or on translation. (b) Regulated genetic device through a feedback controller. (c) Regulated genetic device through a feedforward controller. (d) Output temporal response to a disturbance input. Red shows the response of the unregulated device while blue is the response of the regulated device. \label{fig:3}}
\end{figure*}
A variety of  biomolecular controllers have been implemented  for making the output of a genetic module robust to disturbances on transcription and/or translation (Fig. \ref{fig:3}a).  Disturbances on transcription encompass the  contribution  of the endogenous GRN, variability of the level of transcriptional resources,  and variability in the copy number of the DNA encoding the genetic module, as described in the previous section. Disturbances on translation include variability in translational resources, chiefly the ribosome. A review of  biomolecular controller architectures with a description of the perturbations that can be attenuated by these structures can be found in \cite{BarajasReview} and in \cite{Shakiba2021}. At a high-level, the biomolecular control architectures that have been implemented to date to make the output of a genetic module robust to these perturbations can be grouped into feedback architectures (Fig. \ref{fig:3}b) \cite{huang2018quasi,Jones2022,Aoki:2019aa} and feedforward architectures (Fig. \ref{fig:3}c) \cite{Jones2020,KhammashRes,Ukjin,Bleris:2011aa,VoigtSontag}.  In most cases, the objective has been perfect adaptation to the disturbance or equivalently disturbance rejection (Fig. \ref{fig:3}d). For this reason, biomolecular implementations of integral feedback control have been extensively researched. The next two subsections describe the design and implementation of (integral) feedback controllers and of feedforward controllers, which, under suitable assumptions, can also lead to perfect adaptation.

%overview of: feedback vs feedbforward controllers for distrubance rejection

    \subsection{Feedback controllers}\label{sec:integral}
Feedback control architectures for attenuating the effect of  disturbances on the  output of a genetic module (Fig. \ref{fig:3}b) have been built mostly in bacterial cells  (see, for example, \cite{Aoki:2019aa,huang2018quasi,Nilgiriwala:2015aa})   and more recently are appearing  in mammalian cells  \cite{Jones2022,Frei2022}.     Here, we focus the description on a mammalian controller implementation proposed for achieving integral feedback control \cite{Jones2022}.  
    
 Letting  $x= (m,X)$ and $\alpha(u)$  the transcriptional regulatory function in which $u$ is a control input, the plant dynamics can be written in the general form:
 \[ \dot x =f(x,u,d),\;\; y= h(x),\]
 with $f$ and $h$ implicitly defined.  For integral feedback, the input $u$ should ideally be the integral of the error between a desired reference value of $y$, called $v$, and $y$ itself, that is,  $u=g\cdot \int_0^t(K\cdot v- y(\tau))d\tau$ for some $K>0$ and $g>0$,  or alternatively $\frac{du}{dt}=g\cdot (K\cdot v-y(t))$. The  critical question  that arises is how this control law can be realized via biomolecular processes. A major difficulty is the fact that  if $u$ is the concentration of a molecule in the cell, the integral control law implies that this molecule should not decay otherwise there would be an additional ``$-\gamma\cdot u$'' term in the control law, yielding 
 $$\frac{du}{dt}=g\cdot (v-y(t))-\gamma\cdot u,$$
 which is a ``leaky'' integrator. In practice, any molecule in the cell decays even in the absence of degradation due to molecule dilution arising from cell growth and division. As a consequence, any integrator implemented in the cell will be a leaky integrator.  To overcome this problem, it was proposed to make the integrator faster compared to the dilution rate by letting $g=\bar g/\varepsilon$ with $\varepsilon>0$ small:
  $$ \dot x =f(x,u,d),\;\; y= h(x),\;\;\frac{du}{dt}=\frac{\bar g}{\varepsilon}\cdot (K\cdot v-y(t))-\gamma\cdot u,$$
 a form referred to as quasi-integral control (QIC)  \cite{qian2018realizing}. Assuming $v$ and $d$ to be constant and under stability assumptions on the closed loop system, at steady state it must be the case that $ \bar g \cdot (K\cdot v-y)-\varepsilon\cdot \gamma\cdot u=0$, which as $\varepsilon \rightarrow 0$ implies  that $y= K\cdot v$ independent of $d$. When $\varepsilon$ is small, it is possible to show that at steady state   $y= K\cdot v+O(\varepsilon)$  \cite{qian2018realizing}.

\begin{figure*}[t!] 
\includegraphics[width=1\textwidth]{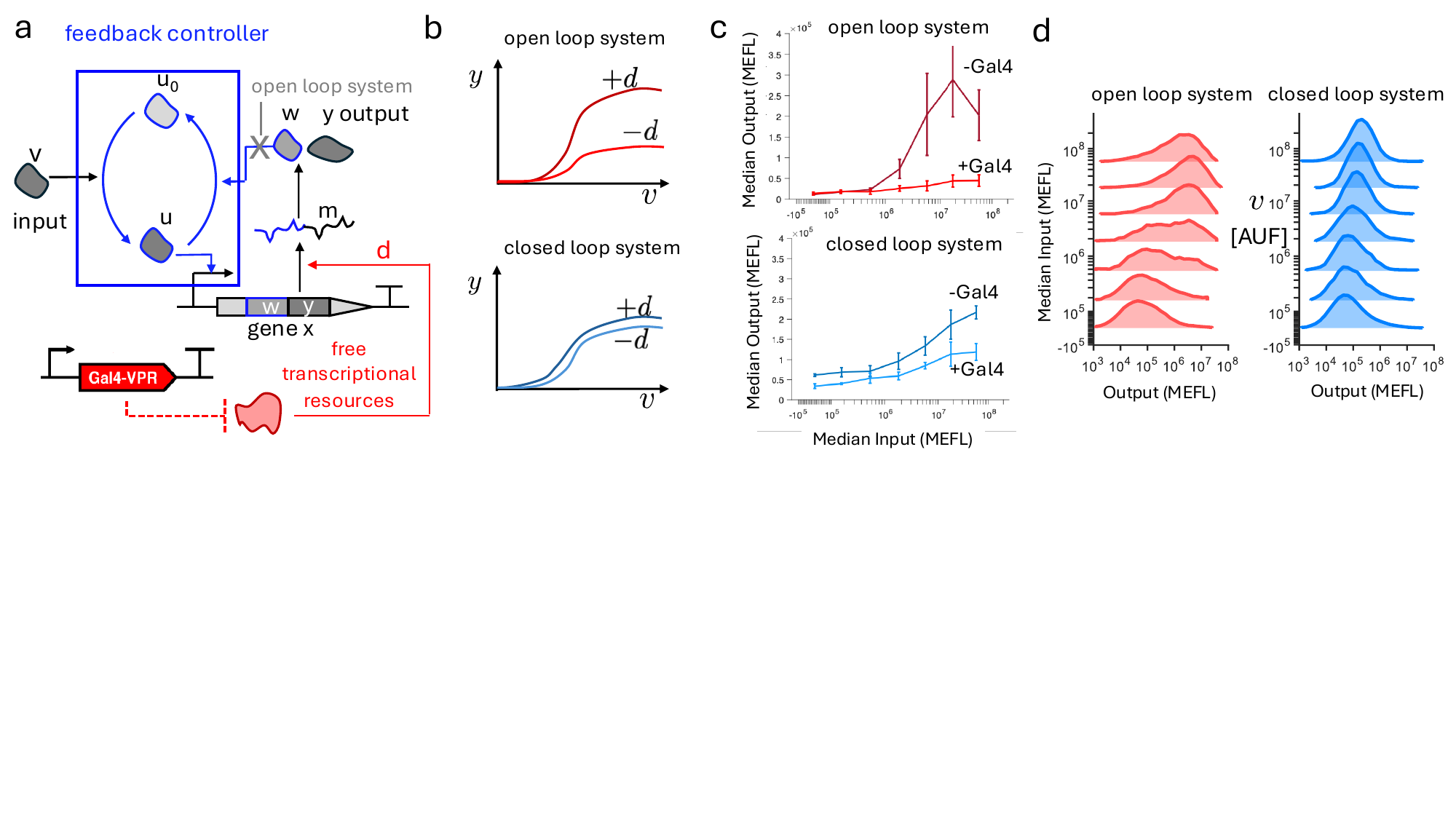}
\vspace{-5cm}
\caption{Biomolecular implementation of an integral feedback cotnroller. (a) Biomolecular circuit diagram. Gene x co-expresses both the output protein and a phosphatase w that dephosphorylates protein u. Only when phosphorylated, this protein can bind to the gene x promoter to activate transcription. Kinase v converts unphosphorylated protein u$_0$ to its phosphorylated version. The open loop system is obtained when phosphotase w is substituted with an inert protein incapable of dephosphorylating u. In the experimental realization published in \cite{Jones2022}, u$_0$ is protein OmpR,    z in the EnvZ kinase, and w is EnvZ mutated to be a pure phosphotase. All these proteins are taken from bacterial cells. OmpR, only when phosphorylated can bind to its cognate promoter cloned in front of gene x.  To activate transcription in mammalian cells, OmpR was fused to an activation domain. (b) Expected behavior of open loop and closed loop systems when the disturbance $d$ in the form of a perturbation to transcription is added. (c) Experimental data showing the output protein level, measured via fluorescence in arbitrary units of fluorescence (AUF), for open loop and closed loop systems when an activator protein (Gal4) is added to the system, which sequesters transcriptional resources, thereby lowering transcription rate. Open loop and closed loop system output levels were made comparable by increasing the DNA copy number in the closed loop compared to the open loop. (d) Distribution of the output $y$ across a cell population for different values of the input for open loop (left) and closed loop (right) systems.   Figures c and d were adapted from  \cite{Jones2022}. \label{fig:4}}
\end{figure*}

 With this solution, the question that remains is to determine what biomolecular process that is much faster than molecule decay can compute the difference between the concentration of two molecules ($v$ and $y$) and integrate it in time. The timescale separation requirement restricts the set of candidate biomolecular processes to those that are faster than the dynamics of gene expression. These include reversible binding (fastest), enzymatic reactions and covalent modifications (faster), and RNA-based reactions (sufficiently fast) \cite{BFS}. Covalent modification cycles, in particular, also called futile cycles, are composed of two opposing enzymatic reactions, in which one transforms a substrate from inactive to active form and the other reverts this. In particular, the dynamics of these enzymatic reactions follow the so-called Michaelis-Menten kinetics \cite{Goldbeter,BFS}, which have two extreme operating regimes: the first order regime and the zero order regime. In the zero order regime, the system enables pure integration as described next. Let v be an enzyme that modifies a substrate protein u$_0$ into an active form u, which then can act as a transcriptional regulator. It is very common for a protein to act as a transcription regulator only after being covalently modified through phosphorylation or methylation. In this case, u is chosen to be such a modified protein. Let then w be a second enzyme that transforms back u into its unmodified form u$_0$. Michaelis-Menten kinetics describe the rate of change of $u$ as follows:
 \begin{equation}
 \frac{du}{dt}=k_1\cdot v\cdot \frac{u_0}{u_0+K_1}-k_2 \cdot w\cdot \frac{u}{u+K_2}-\gamma\cdot u,\\ 
 \label{eq:MM}
 \end{equation} 
 in which $k_1$ and $k_2$ are the catalytic constants of the two enzymatic reactions, $K_1$ and $K_2$ are the Michaelis-Menten constants of the two enzymatic reactions. For the derivation of these dynamical model from the  enzymatic reactions, the reader is referred to \cite{BFS} (Chapter 2). Time scale separation between enzymatic reactions and decay implies the existence of $\varepsilon\ll 1$ such that $\varepsilon=\gamma/k_2$. Furthermore, in the zero-order regime of the enzymatic reactions, it holds that $u_0\gg K_1$ and $u\gg K_2$, leading to the following dynamics
 \[ \frac{du}{dt} =\frac{\gamma}{\varepsilon}\left( \frac{k_1}{k_2}v-w \right)-\gamma\cdot u.\] 
 By letting $w=y$, we obtain the following QIC system:
 \begin{equation}
  \dot x =f(x,u,d),\;\; y= h(x),\;\;
   \frac{du}{dt} =\frac{\gamma}{\varepsilon}\left( \frac{k_1}{k_2}v-y \right)-\gamma\cdot u,
 \label{eq:phoQIC}
 \end{equation}
 whose molecular diagram is represented in Fig. \ref{fig:4}a. This system has a unique asymptotically stable equilibrium point as long as $\partial\alpha(u)/\partial u\geq c>0$ and the mRNA dynamics are much faster than protein dynamics, which is typically satisfied in practice \cite{BFS}. For the technical details, the reader is referred to \cite{qian2018realizing}. Therefore, it follows that at steady state $y=  \frac{k_1}{k_2}v+O(\varepsilon)$. As a consequence,  it is expected that the error between $y$ unperturbed (with $d$ equal to a constant nominal value) and $y$ perturbed (with $d$ constant but different from a nominal value) is smaller in the closed loop system when compared to the open loop system ($w$ is a constant and not  varying with $y$). See Fig. \ref{fig:4}b.

 %implementation 

 This biomolecular control architecture was implemented in mammalian cells by taking as a covalent modification cycle a phosphorylation cycle from bacteria in order to minimize possible unwanted interactions with the intra-cellular mammalian environment \cite{Jones2022}. Because the output molecule to be controlled is given, it cannot be set equal to the enzyme w. Therefore, the feedback was implemented by co-expressing with the gene of protein X the gene of protein w such that the level of w mirrors that of X. This construction theoretically allows $w=y$ (Fig. \ref{fig:4}a). Experimentally, two types of perturbations were applied to the system. In one case, a microRNA targeting the mRNA of protein X for degradation was introduced in the system, embodied by a change in $\delta$. In another case, another gene was expressed in high copies, thereby sequestering transcriptional resources from the genetic module and causing a drop in the module's output \cite{Jones2020}.  The experimental data show that, as expected from theory, the closed loop system's output response to the input (kinase) v is significantly less affected by the disturbance than the open loop system (Fig. \ref{fig:4}c). Also, as expected,  the closed loop system reduces the variability of the output due to variable DNA copy number across cells, which is a consequence of using transient transfection of DNA plasmids in mammalian cells (Fig. \ref{fig:4}d).

 %open questions:
 The theoretical analysis of the closed loop system was performed for constant disturbances, that is, for step changes in resource availability, in mRNA decay rate, and in DNA copy number and for constant reference inputs. It remains to be determined what the performance of the system is when disturbances and/or reference inputs are time-varying, which can be encountered in applications where a trajectory needs to be tracked and perturbations vary with time. Initial theoretical analysis was performed for sequestration-based quasi-integral controllers \cite{huang2018quasi} and relied on the time scale separation between input dynamics and controller dynamics, assuming the controller dynamics to be much faster than the input dynamics.  However,  the results apply only to the linearization of the system, while results for the nonlinear system are still lacking \cite{qian2019singular}. Even for the constant input case, global convergence results are also still lacking as the analysis of \cite{qian2018realizing} was also based on the system's linearization. 
 
 % timevarying still unclear, global stability properties still lacking,  
        
%    -overview of all biomolecular feedback control architectures in a figure: 1. high-gain kayzad; 2. QUIC Yili, 3. Mustafa Nature, 4. Ross Jones 2022

    \subsection{Feedforward controllers}
\begin{figure*}[t!] 
\includegraphics[width=1\textwidth]{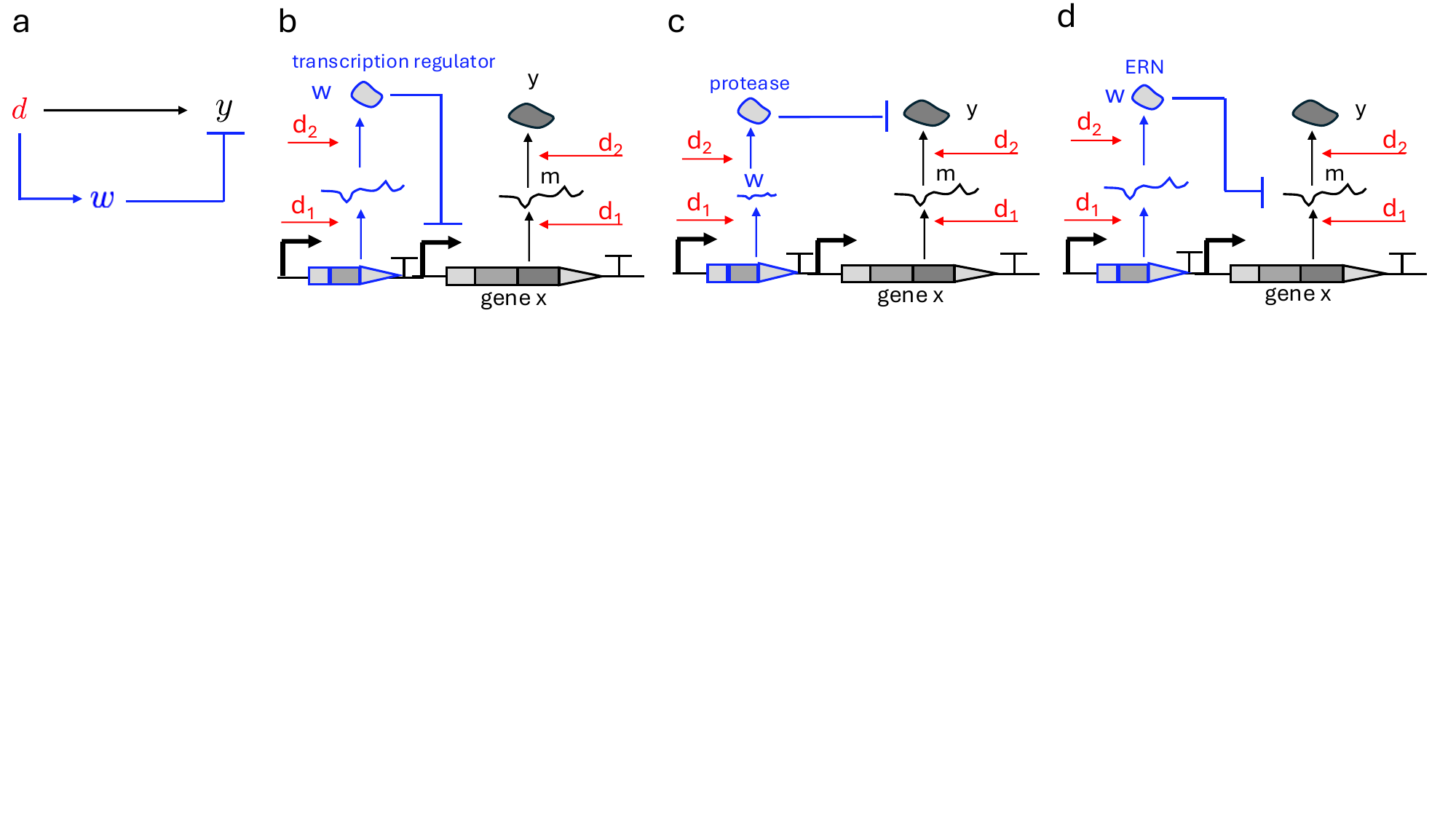}
\vspace{-6.5cm}
\caption{Feedforward biomolecular controllers to attenuate transcriptional and translational perturbations. (a) Incoherent feedforward loop. The disturbance enhances the output $y$ through a direct path and inhibits the output $y$ through an indirect path where $d$ first activates an intermediate variable $w$, which then inhibits $y$. The T-like arrow represents inhibition or repression.  If well tuned, the effects of the two branches on the output can cancel each other. (b) Feedforward genetic controller implementing an incoherent feedforward loop from disturbance inputs $d_i$ to output $y$. The inhibition of y by w is achieved through transcriptional repression. (c) Feedforward genetic controller where the inhibition of y by w is achieved by w catalyzing the degradation of y as a protease. (d) Feedforward genetic controller where the inhibition of y by w is achieved by w catalyzing the degradation of the mRNA m as an endoribonuclease (ERN). Here, $d_1$ represents a perturbation of the transcription rate while $d_2$ represents a perturbation of the translation rate.  \label{fig:5}}
\end{figure*}
%
    
    %overview of feedforward controllers for robustness built, which ones bacterial or mammalian, what we focus on here
    
    Feedforward controllers with an objective of making the output of a genetic module robust to perturbations on transcription and translation have been built in both eukaryotic/mammalian cells \cite{Bleris:2011aa,Jones2020,KhammashRes} and bacterial cells \cite{VoigtSontag,Ukjin}. The designs of \cite{Bleris:2011aa,VoigtSontag} aimed at reducing the uncertainty on the output of a genetic module caused by  variability on the DNA copy number, which is well captured by a perturbation on the transcription rate. The designs of \cite{Jones2020,KhammashRes} were concerned with making the output of the module robust to perturbation in transcriptional/translational resources in mammalian cells, and the design of \cite{Ukjin} had the same objective but in bacterial cells. The following description focuses mostly on the design of \cite{Jones2020} and of \cite{Bleris:2011aa}.

    % differential equations of the genetic module with the ERN on top
    
    The design of a feedforward controller hinges on adding a molecule as part of the controller such that its concentration is affected by the disturbance in the same way as the output  and such that it has an opposing effect to that of the disturbance on the output (Fig. \ref{fig:5}a). This type of network architecture is also referred to as incoherent feedforward loop in the systems biology literature \cite{Alon:2006aa}. If the disturbance in question is on transcription and/or translation, such as in the case of shared resources, it is sufficient to express from an exact copy of the promoter expressing gene x a molecule that requires both transcription and translation resources for its production, hence it must be a protein,  and inhibits the expression of gene x in some way. This protein can inhibit expression of gene x transcriptionally, by acting as a transcriptional repressor (Fig. \ref{fig:5}b), translationally, by acting as a protease that directly degrades the output protein (Fig. \ref{fig:5}c), or as an  endoribonuclease (ERN) that enzymatically degrades the mRNA of gene x (Fig. \ref{fig:5}d). Any of these designs can, in principle, achieve almost perfect disturbance rejection under suitable parameter regimes for  specific disturbances. These include distrubances on transcription, such as variability of DNA copy number and changes in transcriptional resource levels, and on translation, such as variability on translational resource levels. The description here follows the design that uses an ERN as an illustration, but the derivation of the model and analysis for the other designs can be carried similarly.

\begin{figure*}[t!] 
\includegraphics[width=1\textwidth]{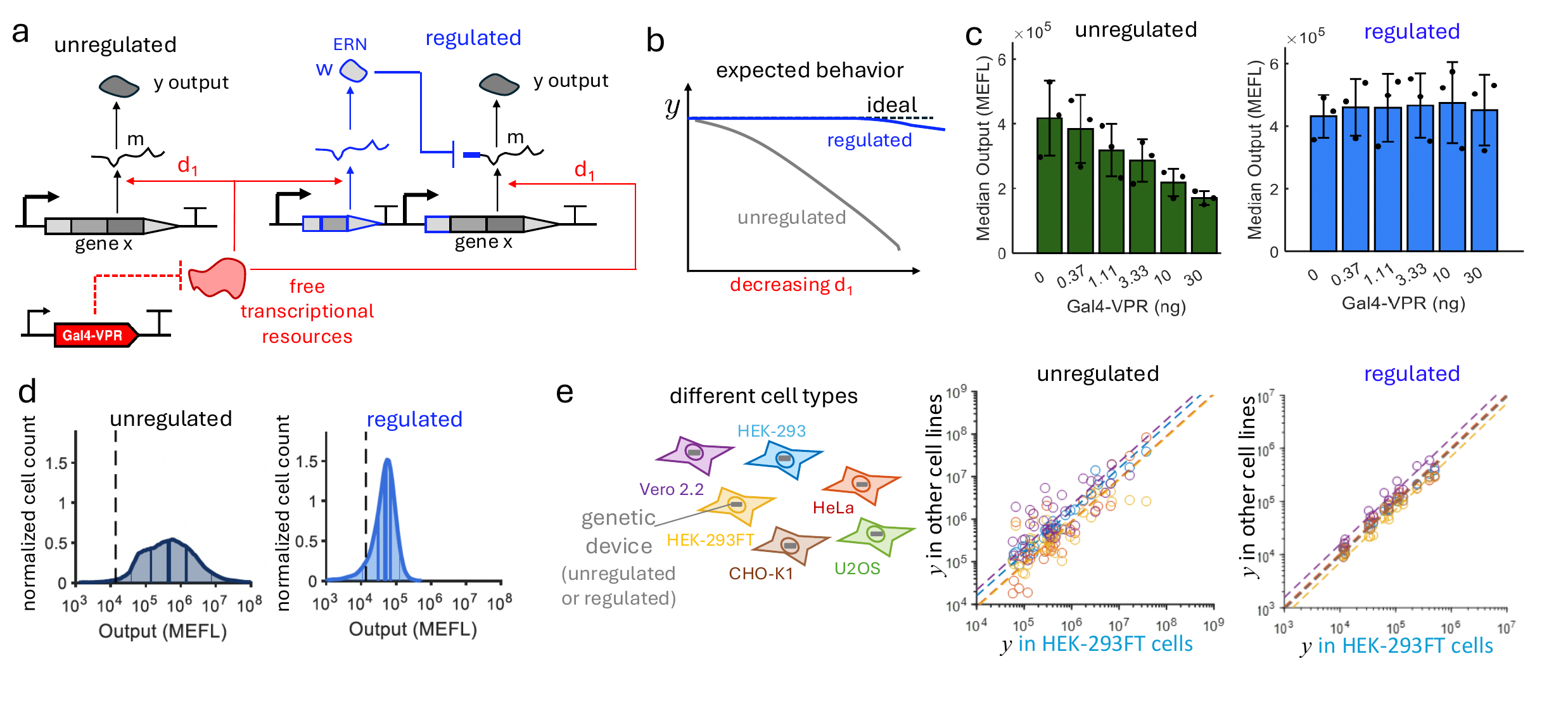}
\vspace{-1cm}
\caption{Implementation of a feedforward controller and its experimental validation. (a) Synthetic genetic circuit implementation of an unregulated genetic device and of its corresponding regulated device via a feedforward controller. The disturbance to the transcription rate is applied by expressing a transcriptional activator (Gal4-VPR), which sequesters transcriptional co-factors from the genetic module, thereby lowering its transcription rate \cite{Jones2020}. (b) Expected behavior of the feedforward controller (in blue). (c) Experimental data showing the output of the genetic module, unregulated or regulated, as the amount of transcriptional activator is increased, thereby lowering the availability of transcriptional resources to the module. The regulated module achieves the same output level as the unregulated one  (with no Gal4-VPR) despite the ERN-based repression by increasing the DNA copy number of the genetic module. (d) Performance of the regulated module with respect to attenuating the effect of variable transcription rate constant due to variability in DNA copy number. (e) Different cell lines have different levels of resources and hence will have different levels of output of the genetic module (unregulated). The regulated device quenches the differences in output due to global changes to resources required for gene expression.  Data plots are taken from \cite{Jones2020}. \label{fig:7}}
\end{figure*}

 For this implementation, the controller dynamics describe the production of the endoribonucleas E from a copy of the promoter expressing gene x, ensuring they both use the same transcriptional resources, which are denoted by $d_1$:
 \begin{equation}
 \frac{dm_E}{dt} = \bar \alpha\cdot {\color{red}d_1} - \bar\delta \cdot m_E,\;\;\;\frac{dE}{dt} = \bar\beta\cdot {\color{red} d_2}\cdot m_E - \bar\gamma \cdot E, 
 \label{eq:ffwd-controller}
 \end{equation}
 in which $d_2$ captures perturbations in the translation rate constant, such as arising from ribosome sequestration.
The plant dynamics with the effect of the controller  are given as follows:
 \begin{equation}
 \frac{dm}{dt} = \alpha\cdot {\color{red}d_1} - \delta \cdot m - {\color{blue}g\cdot m\cdot E},\;\;\;\frac{dX}{dt} = \beta\cdot {\color{red} d_2}\cdot m - \gamma \cdot X, 
 \label{eq:ffwd-plant}
 \end{equation}
 which are obtained considering a one-step reaction model of the enzymatic reaction of the ERN degrading the mRNA, $\ce{m + E ->[g] E}$. It is possible to show that this system has a unique asymptotically stable equilibrium. This equilibrium can be found by setting all the derivatives to zero, to obtain:
 $$E=\frac{\bar\beta\cdot {\color{red} d_2}\cdot { \bar \alpha\cdot {\color{red}d_1} }  }{\bar\gamma\cdot \bar\delta},\;\;\; X=\frac{\beta\cdot {\color{red} d_2}\cdot \alpha\cdot {\color{red}d_1} }{\gamma\cdot(\delta+{\color{blue}g\cdot E})},$$
  and substituting into $E$ its expression, finally leads to the steady state value of $X$ as
  $$X= \frac{\beta\cdot {\color{red} d_2}\cdot \alpha\cdot {\color{red}d_1} }{\gamma\cdot(\delta+{\color{blue}g\cdot \frac{\bar\beta\cdot {\color{red} d_2}\cdot { \bar \alpha\cdot {\color{red}d_1} }  }{\bar\gamma\cdot \bar\delta}})} ,$$
    in which $g$ is a tunable controller parameter via the affinity of the binding site of the RNA on the mRNA and $(\bar\beta\cdot\bar\alpha)/(\bar\gamma\cdot\bar\delta)$ is also tunable by tuning the translation rate constant $\bar\beta$ via the upstream open reading frames (uORFs) in the mRNA \cite{Jones2020}. By calling $\theta=g\cdot (\bar\beta\cdot\bar\alpha)/(\bar\gamma\cdot\bar\delta)$ and $d=d_1\cdot d_2$, the above expression can be re-written as
    $$ X=\frac{\beta\cdot\alpha}{\gamma}\cdot \frac{{\color{red} d}}{\delta+{\color{blue}\theta}\cdot {\color{red} d}}.$$
    As a consequence, by picking $\theta$ sufficiently large such that $\delta\ll \theta\cdot d$, the disturbance disappears from the expression of $X$, which becomes $X\approx (\beta\cdot\alpha)/(\gamma\cdot\theta)$. Because $\theta$ is larger, in order to ensure that the output level in the regulated device is comparable to that of the unregulated device, the value of $\beta$ is increased,  which can be done by varying the uORFs of the X mRNA. The result is that the effect of the disturbances on $y$ can be made smaller  as $\theta$ is increased. In the case in which $\delta=0$, the system exhibits perfect adaptation to the disturbances since the steady state level of the output is completely independent of the disturbance value.
    
    % show how integral control arises and refer to cardinal Alon/Sontag paper
  
       It was shown before that a feedforward controller that exhibits perfect adaptation, under certain technical conditions, has an ``hidden'' integral control action \cite{AlonSontag2011}. In order to discover it, consider the  simplified situation where the only disturbance is $d_1$, so without loss of generality we can set $d_2=1$. Now consider the candidate ``memory'' variable as $z=\frac{E}{q}-\frac{m}{p}$ with $q=(\bar\alpha\cdot\bar\beta)/\bar\delta$ and $p=\alpha$. By taking the derivative with respect to time both sides leads to
       $$\frac{dz}{dt}=\frac{g}{\alpha} E\cdot \left(m-\frac{\bar\gamma\cdot \alpha\cdot\bar\delta}{g\cdot \bar\alpha\cdot\bar\beta}\right)+\frac{\bar\delta}{\bar\alpha}\cdot m_E(t)-d_1,$$
  in which $m_E(t)$ is given by the solution to the differential equation $\frac{d m_E}{dt}=\bar\alpha \cdot d_1-\bar\delta\cdot m_E$. Assuming without loss of generality that $m_E(0)=0$, this differential equation can be solved directly to obtain $m_E(t)=\frac{d_1\cdot \bar\alpha}{\bar\delta} \cdot (1-e^{-\bar\delta\cdot t})$. As a consequence, the $z$ variable satisfied the following differential equation:
  $$\frac{dz}{dt}=     \frac{g}{\alpha} E\cdot \left(m-\frac{\bar\gamma\cdot \alpha\cdot\bar\delta}{g\cdot \bar\alpha\cdot\bar\beta}\right) - e^{-\bar\delta \cdot t},$$
       which, defining the reference constant value of $m$ as $v=\frac{\bar\gamma\cdot \alpha\cdot\bar\delta}{g\cdot \bar\alpha\cdot\bar\beta}$ (independent of the disturbance), is an approximate integrator.  In fact, for $t$ sufficiently large, the exponential vanishes, so the system's steady state must satisfy $m=v$, independent of the disturbance and, as a consequence, also $X=\beta\cdot m/\gamma$ is independent of the disturbance.

    % show experimental results : distrubance attenuation and copy number effect attenuation - comment it is the same for miR and refer to Bleris/sontag paper
         
    This feedforward controller was implemented in mammalian cells by the work of \cite{Jones2020}. The system's ability to reject a disturbance was experimentally assessed by progressively adding more of a transcriptional activator to the cell  (Fig. \ref{fig:7}a). A transcriptional activator  sequesters transcriptional resources away from the genetic module to enact  its activating function, thereby leading to a decrease of transcription rate in the genetic module. Specifically, as more of this activator is added to the cell, more resources are sequestered and, as a consequence the transcription rate $\alpha\cdot d_1$ decreases by virtue of a decreasing $d_1$. The expectation is that as the transcription rate is decreased by increasing the amount of activator, the regulated system should keep a constant output while the unregulated system should experience a decreasing output level (Fig. \ref{fig:7}b). The output response of the regulated system as a function of the activator level was compared to that of an unregulated system with comparable nominal value of the output in the absence of the disturbance. It is critical to compare unregulated and regulated systems for similar output levels because the system is non-linear and if the operating point of the state is different between the two systems, any differences in sensitivities to the disturbance may be due to the nonlinearities of the dynamics as opposed to the  effect of the controller. In order to achieve comparable levels of the output between regulated and unregulated systems, the value of $\alpha$ was chosen to be larger in the regulated system by increasing the DNA copy number. The experimental results are shown in Fig. \ref{fig:7}c.
    
    This feedforward controller also equips the regulated system with robustness to variability in DNA copy number, which is captured by a variable value of $\alpha$ across the cell population. This is shown in Fig. \ref{fig:7}d, in which the distribution of the output level is much narrower in the regulated system compared to the unregulated system. Finally, this feedforward controller was tested in many different cell lines, causing dramatically different intra-cellular contexts with different level of resources for gene expression (Fig. \ref{fig:7}e). In fact, while the unregulated modules' output was highly variable across different cell lines, the regulated module's output was substantially consistent across the different cell lines (Fig. \ref{fig:7}e, compare right with left plots), bolstering the increased robustness to environmental uncertainty that a biomolecular feedforward controller provides.  

\begin{floatingfigure}[r]{9cm} 
\includegraphics[width=1.15\textwidth]{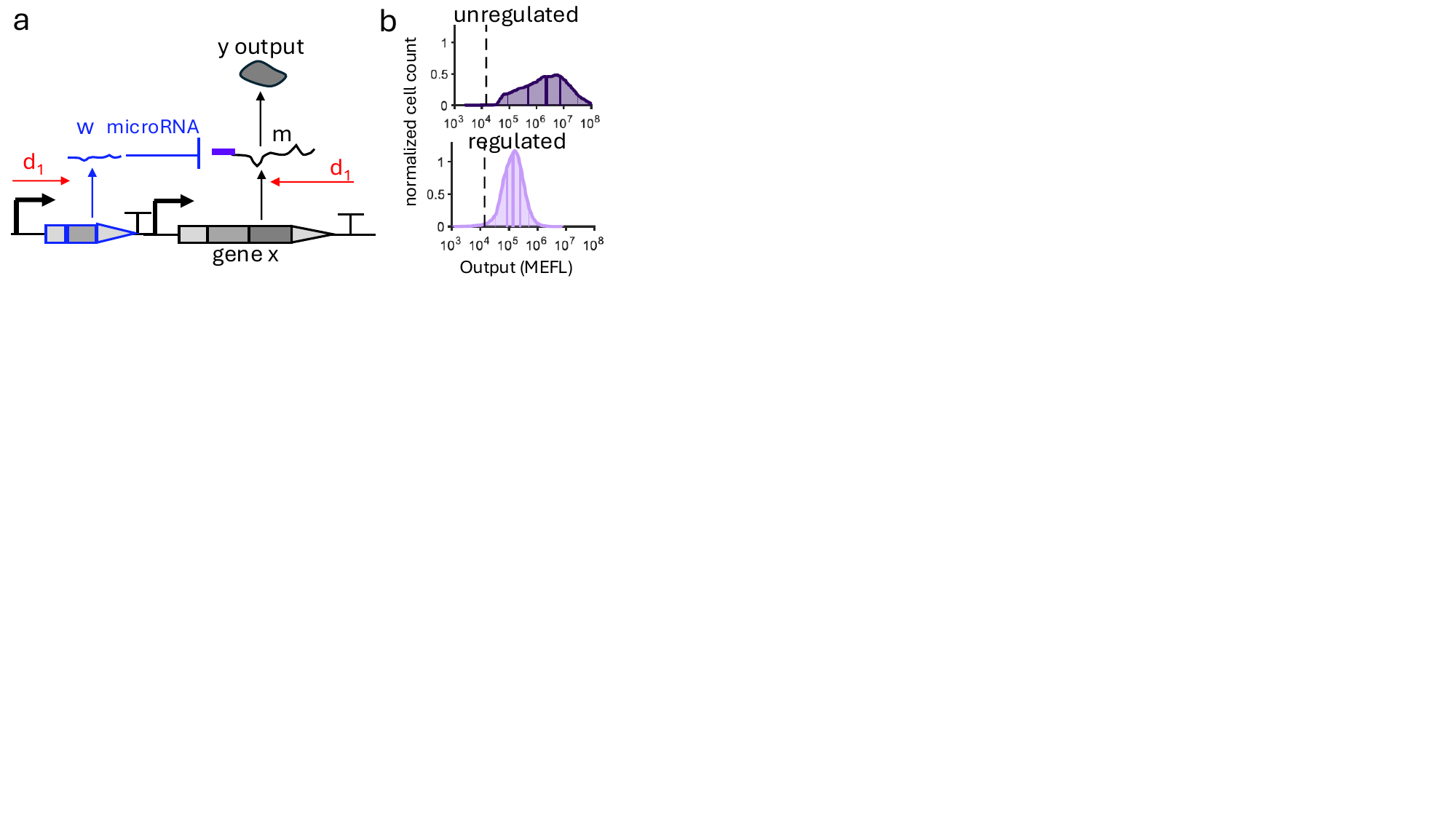}
\vspace{-7.7cm}
\caption{Feedforward controller for attenuating variability in transcription rate constant. (a) Synthetic genetic implementation employing microRNA as the intermediate species w that degrades the output's mRNA by the addition of a microRNA tag on the target mRNA. (b) Distribution of the output level in the unregulated genetic module (no microRNA) and in the regulated geneti module (with microRNA). The latter distribution is significantly narrower than the former.  \label{fig:7b}}
\end{floatingfigure}
%   
    
   % - feedforward control via miR, ERNs and importance of disturbance entry

   A  commonly used implementation of a feedforward controller is one where the controller expresses a microRNA w instead of an endoribonuclease, which enzymatically degrades the mRNA of the genetic module via RNA-RNA interaction (Fig. \ref{fig:7b}a) \cite{Bleris:2011aa}. Because the  production of  w does not require translation, this design will be able to attenuate the effect of transcriptional perturbations but not of translational perturbations. Variation of DNA copy number across a cell population is the source of transcriptional variability that this design is well equipped to attenuate, as seen from the experimental data shown in Fig. \ref{fig:7b}b. To achieve this, it is critical to encode the microRNA on the same piece of DNA as the target mRNA, so that the disturbance $d_1$ due to variable DNA copy number is the same for both. This design, however, is not theoretically guaranteed to attenuate perturbations in the availability of transcriptional resources. This is because in mammalian cells promoters that express microRNAs take different transcriptional resources from those expressing proteins. As a consequence,   addition of an activator protein that sequesters transcriptional resources, may cause a change in the transcription rate of the mRNA but not of the microRNA.  A possible solution to this problem  is to  use the same promoter for the microRNA  as that used for the mRNA, which can be achieved by encoding the short sequence of the microRNA inside an intron of a dummy gene  \cite{Albert}.        
      % comment on remaining open questions: 
      
      Feedforward controllers are widely used in applications as they are easily tunable and require to produce only one  molecule, which, in the case of a microRNA, is also encoded by a very short gene, which is convenient for delivery in the cell. A downside of feedforward controllers for attenuating disturbances on transcriptional resources is that the promoters of the controller molecule w and that of the genetic device need to be the same. This has made it difficult to extend feedforward control design to genetic modules that take a transcriptional regulator as an input, which is desirable in applications where the output of the genetic device should be dynamically modulated. This is still subject of on-going research.

    \section{Case study: Reprogramming somatic cells to pluripotent stem cells}
\begin{figure}[h!] 
\includegraphics[width=0.47\textwidth]{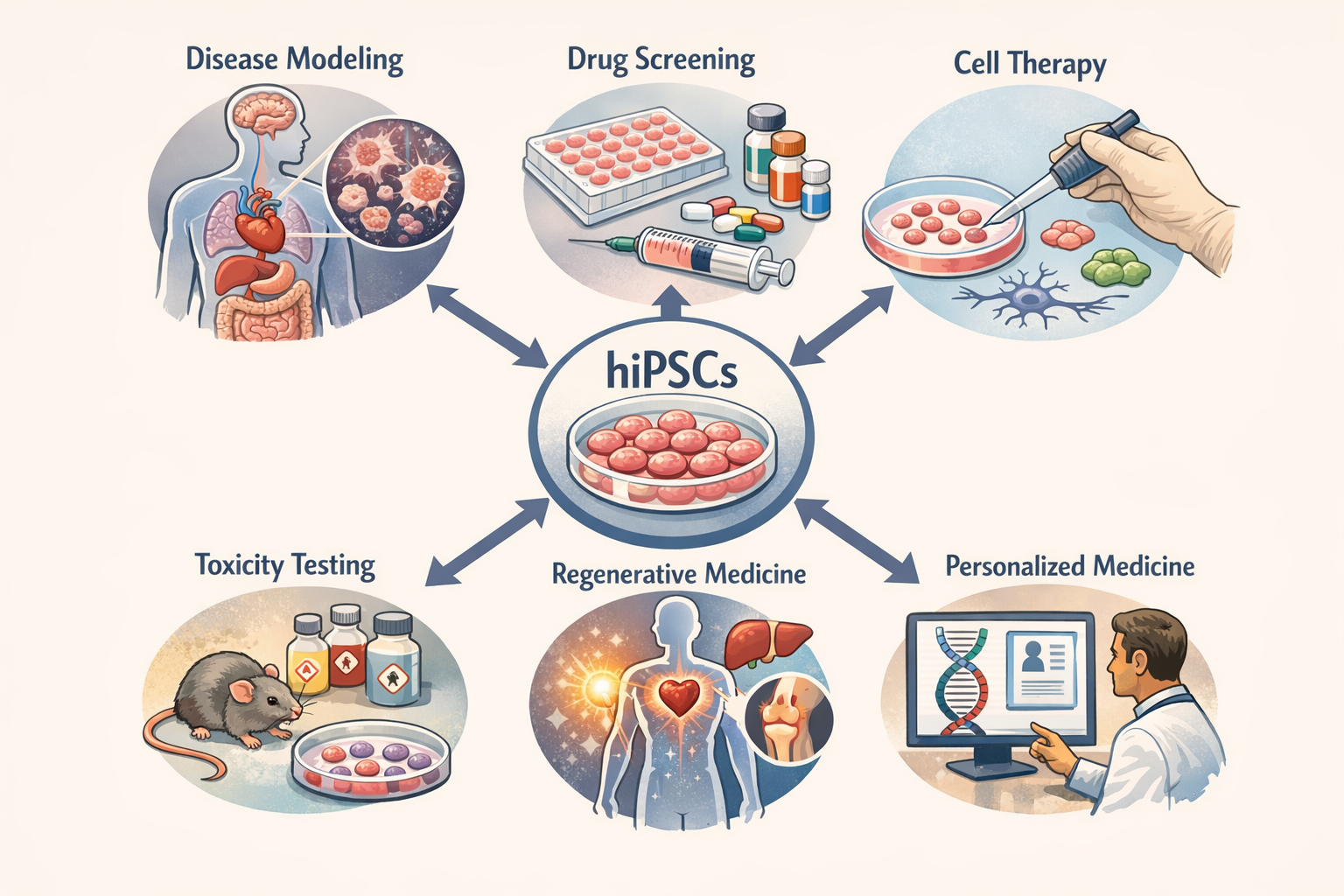}
\caption{Medical relevance of hiPSCs. Image credit: ChatGPT. \label{fig:8}}
\end{figure}
Due to their ability to self-renew and  to differentiate into any cell type,  human induced pluripotent stem cells (hiPSCs)   are revolutionizing regenerative medicine by establishing a self-regenerating bank of patient specific cells that can be differentiated to repair damaged tissues, and are accelerating scientific discovery as a test-bed for drug screening, disease modeling, and toxicity testing (Fig. \ref{fig:8}).  For example, hiPSCs can be engineered to express therapeutic agents, such as   chimeric antigen receptor (CAR) proteins, and then be differentiated to T cells to become potent cancer-killing machines to be delivered to patients with  immunotherapy  \cite{TCell}. Similarly, they can be differentiated in all cell types of the body and, as such, can be used to regenerate damaged  tissues, including brain cells, thereby offering a new way to treat diseases such as Parkinson \cite{GDNF-Parkinson}. Patient specific hiPSCs can be further used to generate  cells for cell therapy in a variety of diseases, including brain and heart disease   \cite{RegenMedicine}.

%background in hiPSC reprogramming: why we want to do it

\subsection{Reprogramming as controlling a dynamical system}
hiPSCs can be derived from somatic cells through the process of  hiPSC reprogramming. Originally, it was  believed that the process of cellular differentiation, from stem cells to the cell types of the body, was unidirectional. The Waddinghton epigenetic landscape \cite{wadding} is a famous metaphor that captures  this unidirectionality, in which differentiation is represented as a ball rolling down a hilly landscape and loosing differentiation potential along the way to end into one of many possible basins representing  different terminally differentiated cell types (Fig. \ref{fig:9}).  It was in 2006 that Yamanaka and colleagues discovered that the process of differentiation and be reversed and that a somatic cell can be reverted to a pluripotent stem cell, although with very low probability \cite{YamanakaReview}. 

The rationale of the reprogramming process is to artificially introduce  pluripotency transcription factors in the cell, including Oct4 and Sox2, which are found at low levels in somatic cells but are highly expressed in pluripotent stem cells. The pluripotency GRN, composed of Oct4, Sox2, and Nanog TFs \cite{Boyer} (Fig. \ref{fig:9}), which are self and mutually activating TFs, is a multi-stable network responsible for maintaining the pluripotent state characterized by high levels of these factors. The reprogramming strategy based on the artificial overexpression of  pluripotency TFs functions by triggering a state transition in this GRN that switches the cell state from a low level of these factors (in somatic cells) to a high level of these factors (in pluripotent stem cells). Once the state transition is triggered, the overexpression can be eliminated as the GRN will be stably maintaining the pluripotent state \cite{FCT2014,CellSystems2017,triad6}. 
\begin{floatingfigure}[r]{9cm} 
\includegraphics[width=0.8\textwidth]{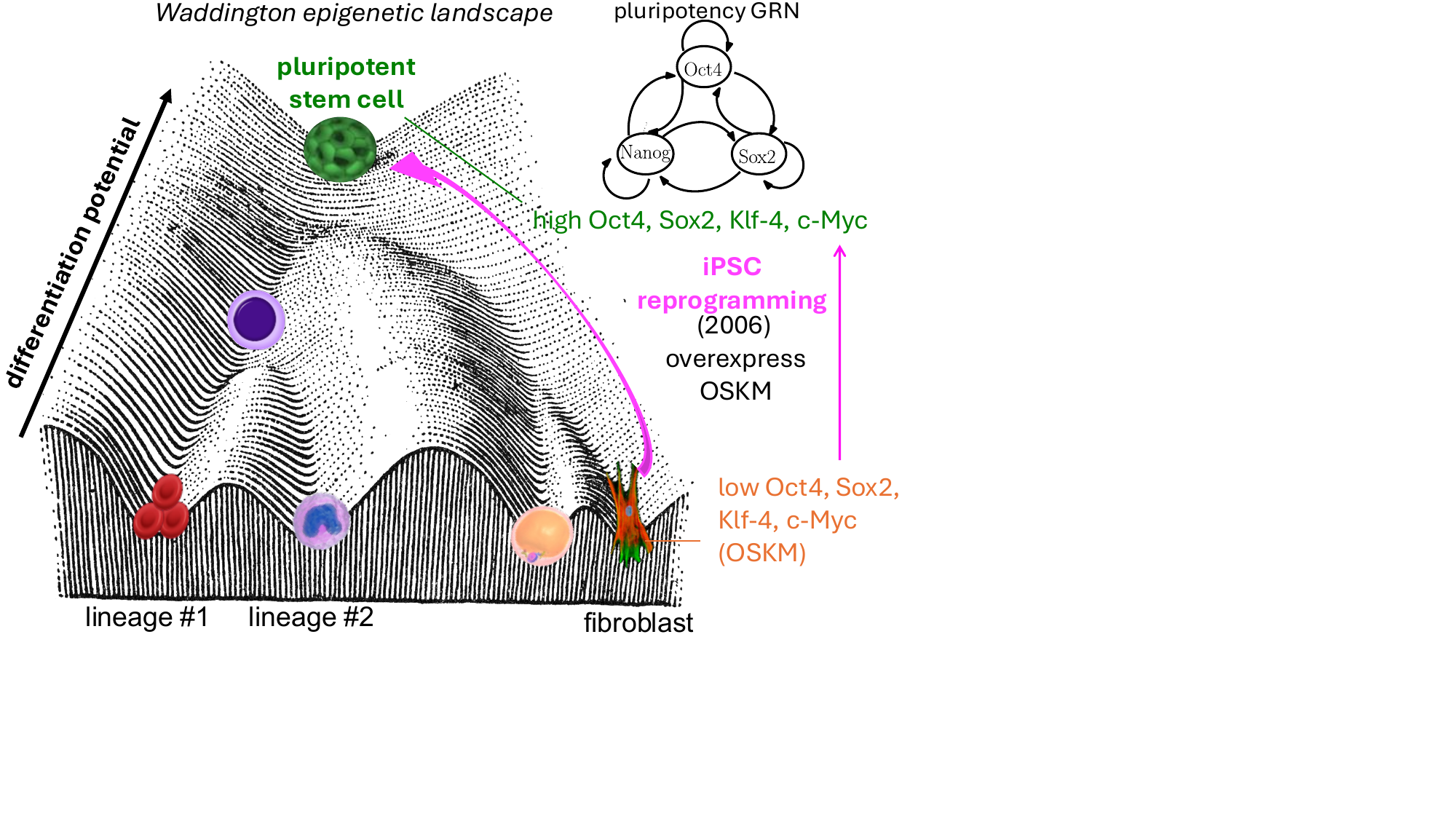}
\vspace{-2.4cm}
\caption{Waddington epigenetic landscape \cite{wadding}. Pluripotent stem cells have the highest differentiation potential since they can differentiate into any cell type. Terminally differentiated cells typically have low levels of pluripotency TFs (Oct4, Sox2, Nanog) while pluripotent stem cells have high-levels of these TFs. iPSC reprogramming consists in artificially expressing pluripotency-inducing TFs in somatic cells \cite{YamanakaReview} in order to reach a self-maintaining state where high levels of these factors are maintained by the pluripotency GRN \cite{Boyer}. \label{fig:9}}
\end{floatingfigure}

It was shown that there is a critical requirement on the level of the Oct4 TF for both entrance into  and maintenance of pluripotency \cite{Niwa2000,Radzisheuskaya2014,Shi2010,papa,Radzisheuskaya2013} and that the level of this factor influences the efficiency of the reprogramming process \cite{papa}, which  remains very low \cite{Schlaeger2015a}. Traditional approaches to hiPSC reprogramming simply introduce DNA that encodes for the TFs to be overexpressed in the cell \cite{YamanakaReview}. These approaches   result in cellular TF concentrations that are highly variable across the cells and are not fully controllable since the endogenous GRN also affects these levels. It is therefore unclear how this indiscriminate overexpression of factors could succeed in reprogramming cells to pluripotency. An instance  of  possible failure modes that could explain the low success rate of iPSC reprogramming can be analyzed by studying the dynamics of the pluripotency GRN under constant inputs, which model constant overexpression of the GRN's transcription factors. 

Specifically, the dynamics of this GRN were analyzed before and   parameter values can be chosen to give a tristable system with the intermediate Oct4/high Nanog combination corresponding to the pluripotent state and low Oct4/low Nanog  corresponding to a somatic cell state (Fig. \ref{fig:10}ab).   Overexpression of Oct4, corresponds to adding a constant production term in the differential equation governing the Oct4 dynamics:
$$\frac{dx_i}{dt}= H(x)-\gamma\cdot x_i+{\color{blue} u_i}$$
in which $H(x)$ represents the influence of the transcriptional regulation of the GRN, $x$ is a vector of transcription factor concentrations, including that of Oct4 as $x_i$, and $u_i$ is the rate of production resulting from artificially introducing the Oct4 gene in the cell. This simple model assumes that the mRNA dynamics are at the quasi-steady state, so the term $H(x)$ is  including  the coefficient $\kappa/\delta$ (see equation (\ref{eq:plant})). Assuming that the initial state of the network has low Nanog and low Oct4 levels, like in the case of a terminally differentiated cell type, overexpression of Oct4, that is, setting $u_i$ to a positive constant will change the number and stability of equilibria and, for very large $u_i$ will turn the network into a monostable system with a unique equilibrium at high Nanog and high Oct4 levels, which does not correspond to the pluripotent state. This is shown by the bifurcation plot of Fig. \ref{fig:10}c. This plot  shows a  configuration of the parameters where a value of $u_i$ does not exist  to switch the state from the low to the intermediate Oct4 state. Indeed, for a monotone network, like the pluripotency network, it is possible to mathematically prove  that its state can be reprogrammed to the high-state by a constant input, but not necessarily to any other state that is not maximal in the component-wise partial ordering of the state space \cite{CellSystems2017}. %see slide 10 of ACC plenary...introduce the triad, the nullclines and the jumping of PL, say it is a monotone net so no guarantee by overexpression the state could be achieved if intermediate.

\begin{floatingfigure}[r]{9cm} 
\includegraphics[width=0.95\textwidth]{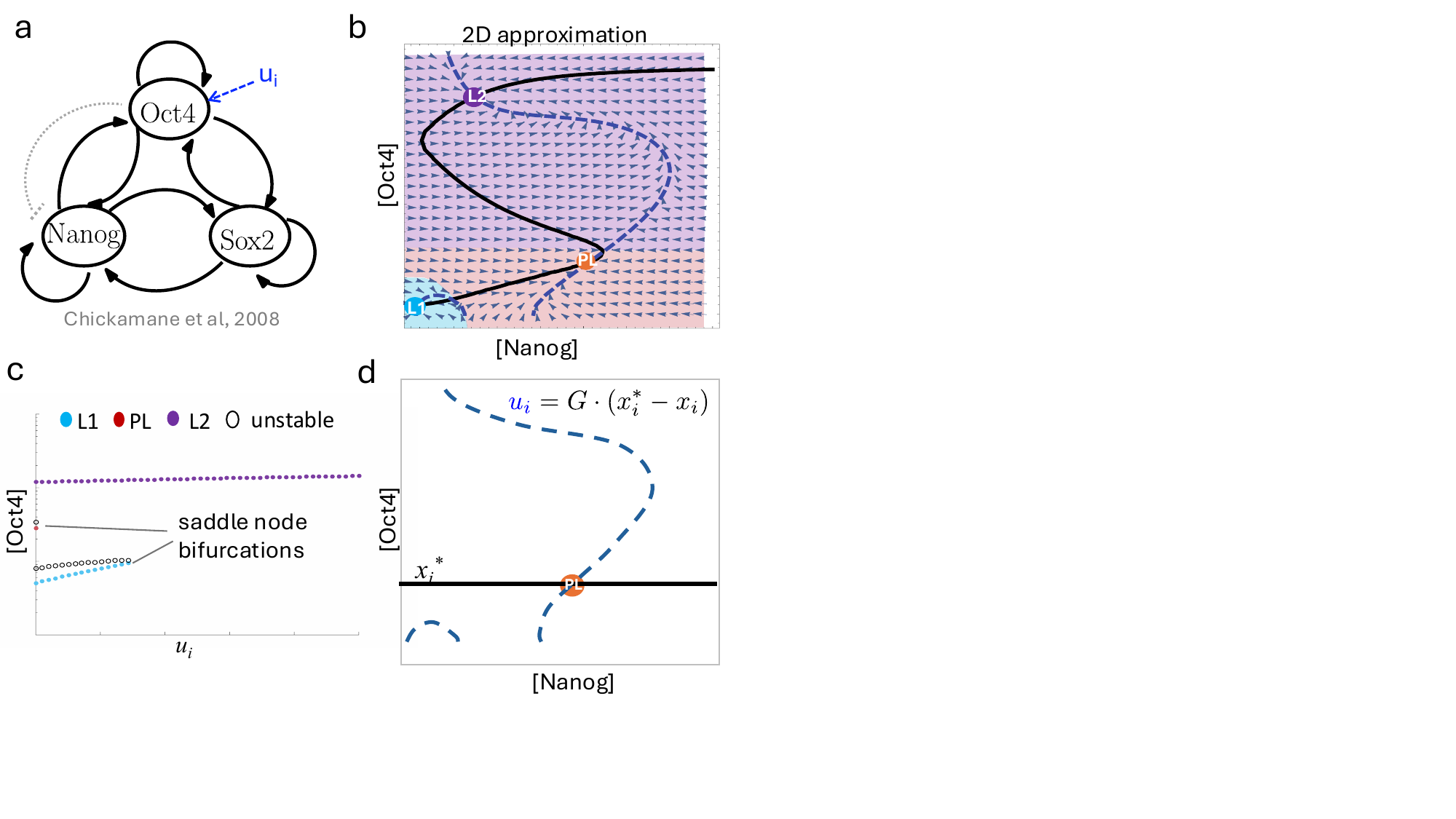}
\vspace{-2cm}
\caption{Reprogramming the pluripotency GRN. (a) The pluripotency GRN discovered in \cite{Boyer} along with an additional repression from Oct4 to Nanog  (T-like arrow) included in the model by \cite{Cdx2Gata62008}. Here, $u_i$ represents over-expression. (b) Nullclines and vector field of a two-dimensional approximation of the GRN in (a), in which Oct4 and Sox2 have been lumped into one state \cite{CellSystems2017}. (c) Bifurcation diagram showing the equilibria and their stability as the overexpression value $u_i$  is varied. Parameter values are in \cite{CellSystems2017}. Here, L1 is a stable steady state representing  a somatic cell phenotype. (d) For $G$ very large, the Oct4 ($x_i$) nullcline becomes a horizontal line with $x_i=x_i^*$. Figures are adapted from  \cite{CellSystems2017}. \label{fig:10}}
\end{floatingfigure}

Given the potentially complicated stability landscape created by the GRN via $H(x)$, it is not possible to rely on a constant input to reprogram the cell to a defined state since this input may not exist and even if it existed it may be not possible to set it with  precision sufficient for reprogramming to the desired target state. A more promising approach is to use feedback to ensure that the desired value of $x_i$ is achieved:
$$\frac{dx_i}{dt}= H(x)-\gamma\cdot x_i+{\color{blue}G\cdot (x_i^*-x_i)},$$
in which $x_i^*$ is close to the value of $x_i$ in the pluripotent state.  If $G>0$ is sufficiently large, the steady state value of $x$ will be close to that of $x_i^*$ (Fig. \ref{fig:10}d). This can be shown mathematically by considering that $H(x)$ is a globally bounded function, which is also positive. Letting $D>0$ be such that $H(x)\leq D$ for all $x$ and $x(0)=0$ without loss of generality, it follows that $\frac{G\cdot x_i^*}{\gamma+G}\left(1-e^{-(\gamma+G)\cdot t}\right) \leq x_i(t) \leq \frac{D+G\cdot x_i^*}{\gamma+G}\left(1-e^{-(\gamma+G)\cdot t}\right)$. As a consequence, $|x_i(t)-x_i^*|$ will become smaller as $G$ increases.

Once the control action is removed, i.e., $u_i=0$, if the state $x$ is  in the region of attraction of the target pluripotent state, then $x(t)$ will converge to it. Thus, if this high-gain feedback strategy is used for all coordinates of $x_i$ and $x^*$ is anywhere in the region of attraction of the target pluripotent state, then the system will be reprogrammed to the target state once the input is removed. In practice, however, it is not possible to control   all TFs composing the GRN, partly because they may not be all known and partly because it is still experimentally difficult to overexpress many TFs concurrently. A  theoretical question of practical relevance that still remains is   how many and which components of $x_i$ need to be controlled   to ensure reprogramming to the target state after the control input is removed, or whether/when a time-varying control over a few $x_i$'s can ensure landing in the region of attraction of the pluripotent state.

 The rest of this section focuses on the accurate control of one TF only as an illustrative example. This is motivated by the fact that    it is possible to reprogram somatic cells to pluripotency by overexpressing Oct4 only, although with very low efficiency  when using  indiscriminate overexpression \cite{Hammachi2012}.

%the importance of Oct4 levels

%cite our repro paper on a defined Oct4 level required through reprogramming - but current approaches to indiscriminate without control

 % \subsection{Perturbations affecting the genetic module during reprogramming}
  
   %describe what the main pertrubations are: H and MOI variability across cells
   
   % describe the effect of H: multi-stability, can give a switch, so quite harmful for precise middle setting
    
    \subsection{Implementation of a feedback/feedforward biomolecular control architecture to achieve a defined level of Oct4}
   This section describes how the high-gain feedback strategy introduced in the previous section was implemented. Although the integral feedback architecture described in the previous sections could be used in principle, in practice, that is guaranteed to achieve perfect adaptation to constant perturbations only. Instead, $H(x)$ is not necessarily a constant perturbation since $x$ dynamically changes during the reprogramming process. However,  it is still bounded, given that the Hill functions are saturating functions \cite{BFS}. Therefore, a high-gain feedback strategy is more promising. To implement a high gain feedback control, it is necessary to introduce genetic constructs in the DNA of the cell that achieve a large production rate of $x_i$ to realize the $G\cdot x_i^*$ term. Concurrently,  a large degradation of $x_i$ must be enforced to realize the large negative feedback $-G\cdot x_i$, which can be physically interpreted as enhanced degradation. Because the perturbation $H(x)$ is a transcriptional perturbation, it is useful to write the dynamics of this genetic construct by explicitly describing mRNA and protein dynamics as follows:
 $$\frac{dm_i}{dt}={\color{red} H_i(x)}-\delta \cdot m_i,\;\;\frac{dx_i}{dt} = \kappa\cdot m_i-\gamma\cdot x_i.$$ 
  A large production rate for x$_i$ can be achieved by inserting DNA in high copy expressing the gene of x$_i$, such that the transcription rate will be $G\cdot \alpha$ for some $\alpha>0$ and $G$ large. A large degradation rate can be easily implemented on the mRNA by using a microRNA targeting the  mRNA m$_i$, encompassing both the artificially introduced and the endogenously present mRNA. By letting $\mu$ represent the microRNA, the dynamics of $x_i$ are now described by
 \begin{equation}
 \begin{aligned}
 & \frac{dm_i}{dt}={\color{red} H_i(x)}-\delta \cdot m_i +{\color{blue} G\cdot \alpha}- {\color{blue} c\cdot m_i\cdot \mu},\;\;{\color{blue}\frac{d\mu}{dt}=G\cdot\beta-\delta\cdot \mu},\\
 &\frac{d x_i}{dt}= \kappa\cdot m_i -\gamma\cdot x_i,
 \label{eq:repro}
 \end{aligned}
 \end{equation}  
   in which $\alpha$ and $\beta$ are positive constants and we have assumed for simplicity a one-step reaction model for the enzymatic reaction by which the microRNA degrades the mRNA with rate constant $c$. %$\ce{m + \mu ->[d] \mu}$.
   
\begin{figure}[t!] 
{\hspace*{3.6cm}\includegraphics[width=1\textwidth]{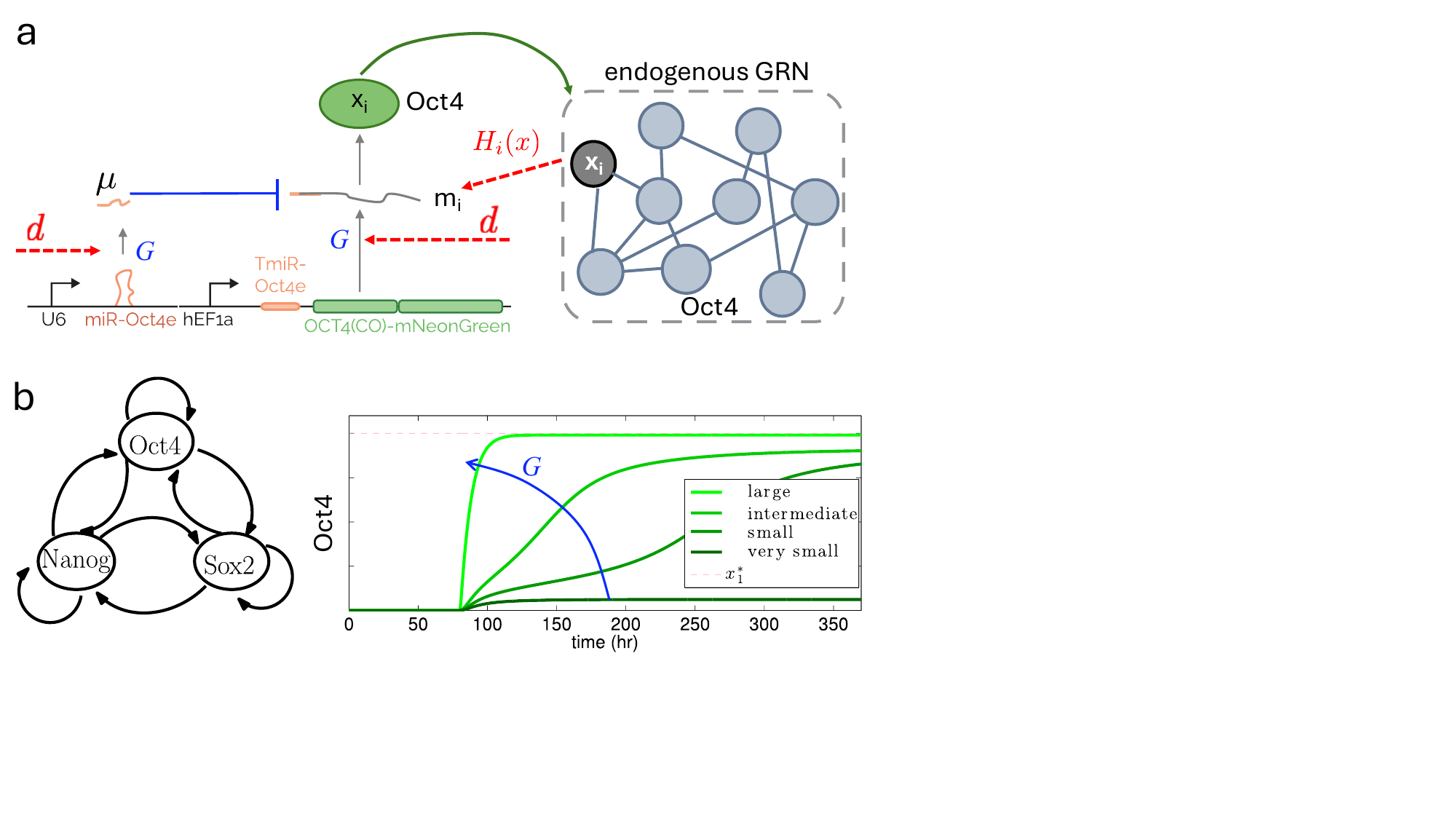}}
\vspace{-2.5cm}
\caption{Synthetic genetic implementation of high-gain feedback and feedforward control.  (a) Oct4 and its microRNA are encoded on the same piece of DNA. Because the DNA is the same, transcriptional disturbances due to variability in DNA copy number $d$  are equal.  The endogenous GRN, including the pluripotency GRN expresses the mRNA of x$_i$ (Oct4), which is captured by a distrubance $H_i(x)$. Both production rate of mRNA and microRNA re large by virtue of high DNA copy number $G$. The microRNA actively degrades the mRNA (indicated by the T-like arrow). (b) Numerical simulations of a system of ordinary differential equations describing the dynamics of the genetic feedback controller in panel (a) and of the pluripotency GRN. Figures are adapted from \cite{CellSystems2017}.  \label{fig:11}}
\end{figure}

   Additionally, a practical consideration pertains to the method used to deliver this DNA   to the cell. Independent of whether lentivirus infection or the so-called piggyBAC transduction are used, the synthetically constructed DNA will be randomly integrated in the chromosome of the cell. The number of copies that are integrated can be to some extent controlled in order of magnitude and the higher the copy number the larger the value of $G$ will be. This number, as explained before, is not easily controlled. This implies that in equations (\ref{eq:repro}) the transcription rates by the synthetic genetic construct will be subject to variability, that is, $G \cdot \alpha$ becomes ${\color{red} d_1} \cdot G\cdot \alpha$ and $G\cdot\beta$ becomes ${\color{red} d_2} \cdot G\cdot \beta$, in which $d_1$ and $d_2$ are the disturbances on the transcription rate of the mRNA and on the microRNA, respectively.
The control strategy should therefore also seek some level of robustness to these variable transcription rates. To this end, a feedforward controller implemented via microRNA such as the one of Fig. \ref{fig:7}a can be employed. Since the high-gain feedback system already uses a microRNA to implement the enhanced degradation term and attenuate the effect of $H(x)$, a feedforward controller of DNA copy number is obtained by placing the microRNA and the gene expressing x$_i$ on the same piece of DNA instead of putting them on different pieces. This ensures that the transcription of the microRNA and that of the mRNA of x$_i$ are subject to the same transcription rate disturbance. Therefore, $d_1=d_2=:d$ and equations (\ref{eq:repro}) can be rewritten as
 \begin{equation}
 \begin{aligned}
 & \frac{dm_i}{dt}={\color{red} H_i(x)}-\delta \cdot m_i +{\color{red} d}\cdot {\color{blue} G\cdot \alpha}- {\color{blue} c\cdot m_i\cdot \mu},\\
 &{\color{blue}\frac{d\mu}{dt}={\color{red} d}\cdot G\cdot\beta-\bar\delta\cdot \mu},\\
 &\frac{d x_i}{dt}= \kappa\cdot m_i -\gamma\cdot x_i.
 \label{eq:repro1}
 \end{aligned}
 \end{equation}  
   The genetic diagram of this feedback-feedforward controller is represented in Fig. \ref{fig:11}a
 To understand how high-gain negative feedback emerges from this system, assume the microRNA dynamics is at its  steady state $\mu=({\color{red}d}\cdot G\cdot \beta)/\bar\delta$  and consider the mRNA dynamics only:
 \begin{equation}
 \frac{dm_i}{dt}={\color{red} H_i(x)} +{\color{blue} G}\cdot {\color{red} d}\cdot \left(\alpha-\frac{c\cdot \beta}{\bar\delta}\cdot m_i\right)-\delta\cdot m_i.
 \label{eq:mRNA}
 \end{equation}
Here, the high-gain feedback term is given by  ${\color{blue} G}\cdot {\color{red} d}\cdot \left(\alpha-\frac{c\cdot \beta}{\bar\delta}\cdot m_i\right)$, while the feedforward action with respect to disturbance $d$ is manifested by the fact that the whole high-gain feedback term is multiplied by $d$. The feedback-feedforward control system (\ref{eq:repro}), for each constant value of $H(x)$, has a unique asymptotically stable equilibrium point given by
 \[m_i=\frac{{\color{red} H_i(x)}+{\color{blue} G}\cdot\alpha\cdot {\color{red}d}}{\delta +c\frac{{\color{blue}G}\cdot\beta\cdot{\color{red} d}}{\bar\delta}}.\] 
From this expression, it follows that as $G\rightarrow\infty$ the mRNA level will be independent of both $H(x)$ and of $d$:
$m_i\rightarrow \frac{\alpha \cdot \bar\delta}{c\cdot \beta}$, which leads to 
$$x_i\rightarrow \frac{\kappa\cdot \alpha \cdot \bar\delta}{c\cdot \beta},$$
   which is also independent of the perturbations. In practice, $H(x)$ will not be a constant, therefore the above derivation needs to be changed as follows. For $G$ sufficiently large, the term $\delta\cdot m_i$ can be neglected in equation (\ref{eq:mRNA}), and letting $D>0$ be a constant such that $H_i(x)<D$ for all $x$, the system is linear  with a bounded time-varying disturbance and, therefore, it follows that  for $t\rightarrow\infty$ the mRNA concentration satisfies 
   $$m_i(t)\rightarrow \frac{\alpha }{\frac{\beta\cdot{ c}}{\bar\delta}}+\tilde m_i,$$
   in which
   $$\tilde m_i\leq \frac{{\color{red} D}}{c\frac{{\color{blue}G}\cdot\beta\cdot{\color{red} d}}{\bar\delta}},$$
   which can be made arbitrarily small by increasing $G$. As a consequence, for $t\rightarrow \infty$, the value of $x_i$ satisfies
   $$x_i(t)\rightarrow  \frac{\kappa\cdot \alpha }{\gamma\cdot \frac{\beta\cdot{ c}}{\bar\delta}}+\tilde x_i,\;\;\; \tilde x_i \leq \frac{\kappa \cdot {\color{red} D}}{\gamma\cdot c\frac{{\color{blue}G}\cdot\beta\cdot{\color{red} d}}{\bar\delta}},
   $$
   which is confirmed by numerical simulations (Fig. \ref{fig:11}b).
   The nominal level of $x_i$ can be modulated by tuning the value of parameter $c$, which can be decreased by decreasing the binding affinity of the microRNA to the mRNA by adding mismatches in the base-pairing between the mRNA and microRNA sequence. This is an experimentally simple way of tuning the level, which was   used in the experimental testbed described next.  
   %
%   Analysis of the dynamics - highlight the feedback and feedforward
   
  % write the steady state and see behavior with G and with d to tune the level
   
   %show the simulations
   
 Numerical simulations  were performed for an ODE system including the dynamics of the pluripotency network and the biomolecular controller in equations (\ref{eq:repro}) to control the level of the Oct4 transcription factor. For a set value of $x_i^*$ and starting from $x_i(0)\approx 0$,  simulation results show that $x_i(t)$ approaches $x_i^*$ as $t\rightarrow\infty$ for sufficiently high levels of $G$ (Fig. \ref{fig:11}b).

   %describe what control strategies: not phospho because H is possibly time-verying and also large gargo --> high-gain feedback: large production and large degradation
   
   % how to make this expression robust to copy number variation? encode the miR on the same DNA piece as Oct4 so that we have in iFFL to defeat DNA copy number variability
   
   % show simulation results from 2017 paper, deterministic and stochastic

\subsection{Experimental testing of the controller  during hiPSC reprogramming}  
  The combined feedback and feedforward biomolecular controller introduced in the previous section was implemented and delivered lentivirally to fibroblasts \cite{ReproPaper,ReprobioRxiv}. The details of the specific delivery method and the experimental protocol to achieve reprogramming to stem cells can be found in \cite{ReproPaper}. In short, lentiviruses encoding the biomolecular controller depicted in Fig. \ref{fig:11}a with x$_i$=Oct4 were used to insert the synthetic DNA into the   genome of  fibroblast cells concurrently with another lentivirus encoding the additional three critical factors to achieve pluripotency, Sox2, KLF4, and c-Myc \cite{yama2006}. These were encoded using constant overexpression without control.  After insertion of the synthetic genetic constructs into the cell's DNA, cells were put under a 19-day long protocol using pluripotency-promoting growth  media.   On day 19, microscopy analysis was performed to determine the number of generated colonies and cells were analyzed using antibody staining for typical markers of pluripotency to provide a quantitative assessment of the yield of the reprogramming process. 
\begin{figure}[h!] 
\hspace*{3 cm}\includegraphics[width=1.1\textwidth]{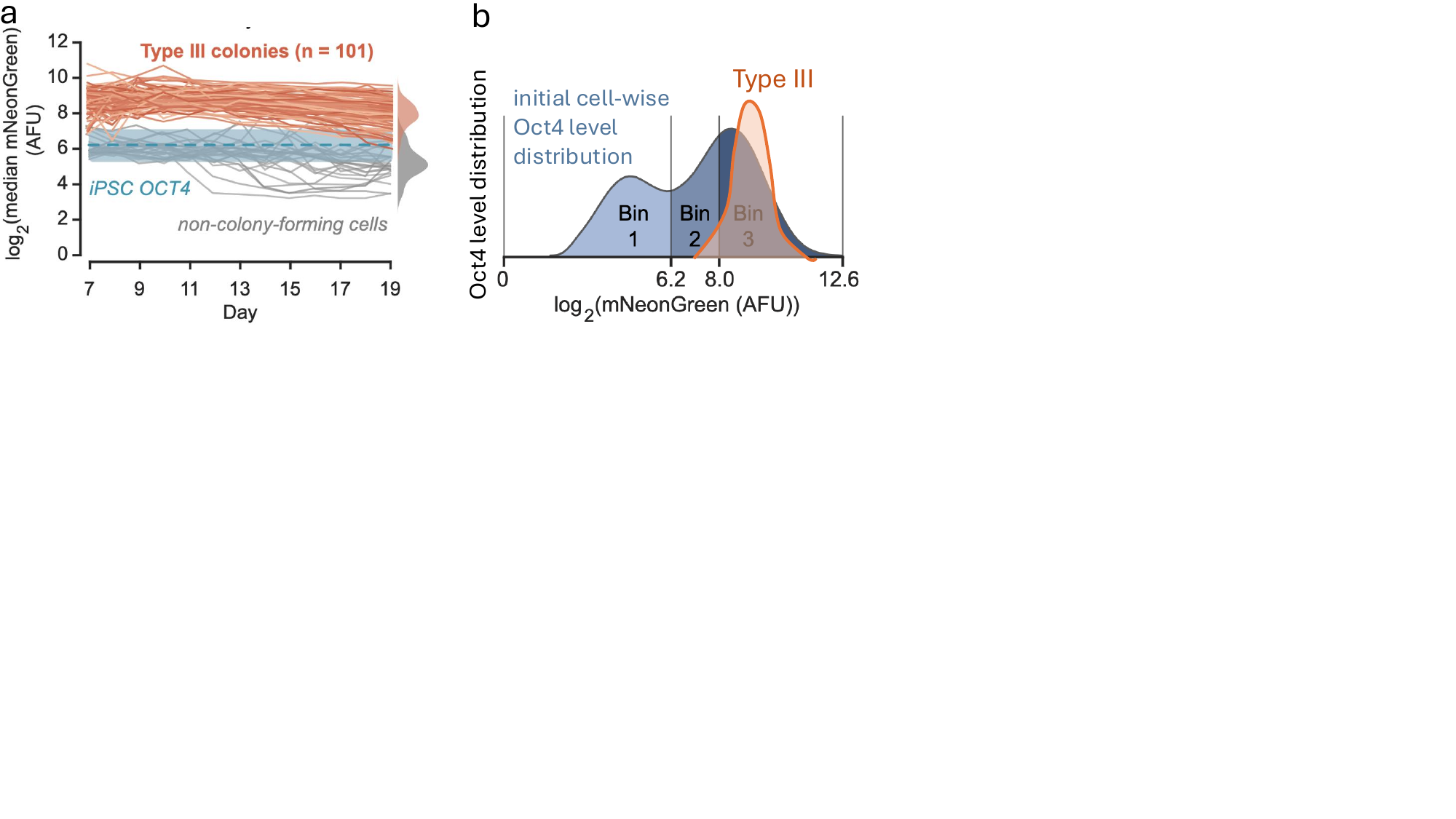}
\vspace{-7cm}
\caption{Desired Oct4 levels for reprogramming fibroblasts to iPSCs.  (a) Temporal trajectories of the Oct4 level in single colonies for those cells that successfully reprogram to iPSCs (orange) and those that fail (gray). Oct4 level was quantified by fusing a green fluorescent protein to the artificially expressed Oct4, while knocking down the endogenous oct4 mRNA \cite{ReproPaper}. (b) Day 19 Oct4 level distribution for Type III colonies compared to the initial oct4 level distribution across the fibroblast cell population. The genetic construct used a traditional inducible promoter to express Oct4, with the addition of a microRNA that would target endogenous Oct4 protein only \cite{ReproPaper}. This ensured that the measured fluorescence was a proxy for the total cellular Oct4 level. Figure was adapted from  \cite{ReproPaper}. \label{fig:12}}
\end{figure}

 Before starting the reprogramming experiments, preliminary experiments were performed to determine the optimal level of Oct4, that is, $x_i^*$, which would be encoded in the synthetic genetic controller by a suitable choice of parameter $c$. These experiments were carried by reprogramming fibroblasts to hiPSCs using a synthetic genetic construct expressing a broad distribution of Oct4 levels, monitoring temporal cell trajectories of Oct4 cellular levels using microscopy through the 19-day timecourse, and by determining what levels/trajectories of Oct4 were enriched within those cells that successfully reprogrammed to hiPSCs (Fig. \ref{fig:12}a). A broad distribution of Oct4 levels was achieved by using an inducible  promoter, widely used in reprogramming experiments, encoded in low copy DNA,  leveraging the fact that reactions at low   molecular counts  increase the variability of molecule concentrations \cite{Swain:2002ww,Raser:2005aa}. The reader is referred to \cite{ReproPaper} for details. The experiment revealed that  successfully reprogrammed cells maintain a defined Oct4 level throughout the entire  reprogramming timeline. The distribution of the Oct4 level across this ``winning'' cell population is very narrow, especially when compared to the broad distribution across the starting cell population that is achieved by the traditional synthetic genetic expression system (Fig. \ref{fig:12}b).  
\begin{figure*}[t!] 
\includegraphics[width=1.1\textwidth]{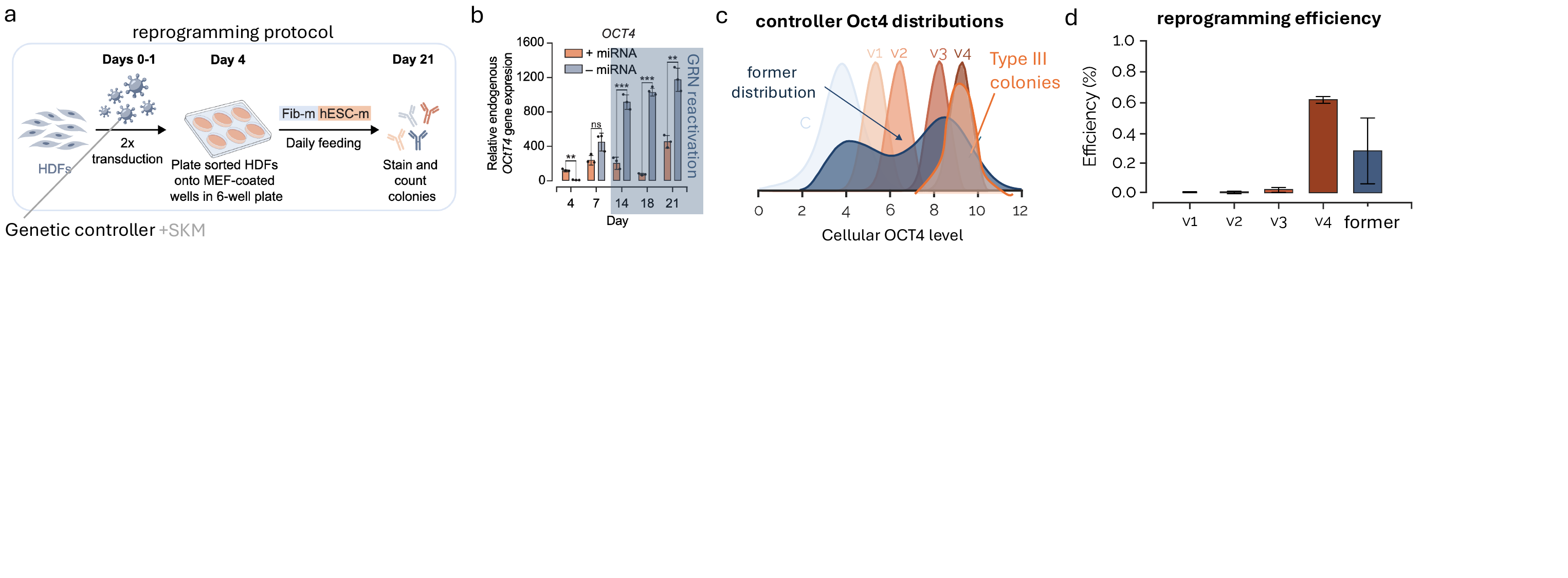}
\vspace{-4.5cm}
\caption{Reprogramming fibroblasts to iPSCs with the controller. (a) The reprogramming protocol starts with the infection fo fibroblasts (HDFs) with viruses containing the genetic controller plus constructs that constitutively express Sox2, KLF4 and c-Myc (SKM). Oct4 endogenously produced mRNA's level is measured by q-PCR along the time course of reprogramming. Orange show data when using the traditional overexpression system with the addition of the microRNA targeting endogenous Oct4 but not artificially produced Oct4 and gray shows the data when using the same  system but without microRNA.  (c) Oct4 level distributions for the four controller variants with stronger (v1) or weaker (v4) binding of the microRNA to the mRNA, tuned by parameter $c$. Gray shows the Oct4 level distribution from the traditional expression system.  Orange plot depicts the Type III colonies Oct4 distribution. (d) Reprogramming efficiency achieved with the different controller variants.  Error bars represent one standard deviation of three independent technical replicates. Figures a and b were adapted from  \cite{ReproPaper}.\label{fig:13}}
\end{figure*}

  Therefore, the objective of the  controller is to achieve this narrow distribution of Oct4 levels across the starting cell population with the aim of ensuring that cells keep the level required for reprogramming. To this end, the feedback and feedforward controller (Fig. \ref{fig:11}a) was inserted in the chromosome of a starting fibroblast cell population via lentiviral delivery. To ensure a large gain $G$, a high multiplicity of infection (MOI) was used for the lentiviruses to ensure that a high number of copies of the DNA encoding the controller would be integrated in the cell chromosome. With this high-gain, the expectation is that the influence of the GRN $H_i(x)$ is attenuated. To experimentally demonstrate this, the level of Oct4 mRNA produced by the endogenous GRN over the course of a reprogramming experiment  was measured. The data show that indeed, the presence of the microRNA in the controller substantially decreases the level of the  Oct4 mRNA produced by the GRN when compared to the same expression system lacking the microRNA (Fig. \ref{fig:13}ab).
  
   Four  variants of the controller were implemented with different values of the parameter $c$ for Oct4  level tuning. The resulting Oct4 distributions across the fibroblast cell population are shown in Fig. \ref{fig:13}c against the Oct4 distribution realized by the traditional constant overexpression system. Among the four variants, one of them,  variant v4, appears to be the one with most overlap with the desired Oct4 distribution across the cells. As a consequence, the expectation is that using variant v4 to reprogram the cells to iPSCs would result in more cells reprogrammed than with the other versions and with the traditional expression system.  To test this hypothesis, the four variants were used to reprogram fibroblasts to iPSCs. The results are shown in Figure \ref{fig:13}d. Variant v4 over-performed  the traditional expression system by about two-fold and all the other variants by orders of magnitude. These results serve as a confirmation that the specific level of Oct4 during reprogramming influences the reprogramming outcomes and that enforcing the desired level improves reprogramming efficiency.

   \subsection{Points of improvement}
These results   overall demonstrate that  tight quantitative control of Oct4 level enabled by the biomolecular controller aids reprogramming. Nevertheless, significant points for improvement remain. Specifically,   among all the cells that during the reprogramming process kept the desired Oct4 level only about 30\% of them led to {\it bona fide} (Type III) hiPSCs \cite{ReproPaper}. Therefore, the expectation would be that  30\% of  the  cells  where the desired Oct4 level is maintained during the reprogramming process successfully reprogram to iPSCs. Instead, the overall reprogramming efficiency achieved by the best version v4 is still below 1\%. Further investigation of the temporal dynamics of Oct4 level during reprogramming revealed that the Oct4 level distribution across the cell population drifts towards lower levels of Oct4 . This contradicts the controller's theoretical ability to maintain cellular Oct4 levels at desired values. A plausible explanation comes from data showing that cells that express higher Oct4 levels results in decreased growth rate, so that they divide less often than cells with lower Oct4 levels. At the level of a population, the result is that  a larger fraction of cells in a population will express lower level as time progresses and therefore, the Oct4 cellular distribution shifts to lower levels. This is consistent with the known role of Oct4 in the cell cycle and would require the addition within the biomolecular controller of mechanisms preventing faster growth of cells with lower Oct4 level. 

A plausible way to achieve this would be to artificially link lower Oct4 levels to cell death by activating apoptotic pathways in cells where the oct4 levels drop below a threshold. These population controllers have been widely developed for bacterial cells \cite{HastyPop1,HastyPop2}, and more recently for mammalian cells \cite{ElowitzPop} and would need to be linked here to Oct4 expression. More sophisticated methods include the development of growth rate controllers, with an overall goal of making growth rate more homogenous across the cell population. In this case,  growth rate could be internally measured by, for example, proteins linked to the cell cycle, and actuated via apoptotic pathways or pathways interfering with cell division. More work is required to develop these controllers and make them functional within the context of a complex biological process such as reprogramming cell fate.

    %show broad experimental distribution obtained at low MOI and no iFFL with superimposed the ideal OCT4 level for winning trajectories - show construct low MOI and no iFFL
    
    % show distributions obtained for construct with high MOI and iFFL - show tuning and tight distribution
    
    %show high gain kills endogenous GRN contribution
    
    % show outcomes of reprogramming efficiencies compared to original - refer to bioRxiv.

\section{CONCLUSION}
In summary, control design approaches are required to achieve desired accuracy of cellular  concentrations of factors involved in important biological processes, such as cell fate determination, the immune system, or production of high-value compounds. Applications of mammalian cell engineering, from regenerative medicine to targeted drug delivery and immunotherapy, will highly benefit from an ability to control precisely the temporal concentration of multiple molecules concurrently inside the cell, and  in response to environmental cues. Biomolecular controllers can be implemented via molecular reactions using the tools that synthetic biology has established, and it is plausible that more such tools will be available in the future since the field is progressing rapidly. These new tools will allow the implementation of increasingly sophisticated control laws that can achieve more demanding performance specifications. 

There are already many examples of biomolecular controllers that have been implemented across organisms for achieving accurate output robustly to perturbations, and this paper described several of them. While the performance of these controllers is promising, they have been experimentally tested mostly in highly controlled conditions   in the context of mammalian or bacterial laboratory cell lines or strains. It remains still unexplored the testing of these controllers within more  realistic biological contexts that are closer to field applications, with the case study of iPSC reprogramming considered in this paper being one isolated example. 

More broadly, synthetic genetic circuits are tuned and tested within the boundaries of well defined laboratory conditions and their operation within a field-relevant environment often remains a question mark, with few notable examples. As for the iPSC reprogramming example, cells bearing the synthetic genetic circuits may be going through metabolic or phenotypic changes, and may affect the functioning of the synthetic circuit through indirect effects, such as on growth rate. These challenges will need to be overcome   by a deeper understanding of the application context and  by designs that are more resilient and context-aware. 

Although controller designs often relies on a coarse model of the plant and its environment, such models are often inaccurate and may bear inappropriate assumptions depending on the specific context. The ability of designing controllers and genetic circuits more generally such that they are aware of the context to some extent is a grand challenge in the field. Machine learning tools are starting to be considered as a way to complement  physics-based models and learn features of the context that can be accounted for during circuit design \cite{JJC-DDV-ML}.
It is plausible that future design techniques will incorporate physical information while leveraging machine learning to shed light on hidden connectivity and context effects, which today still remain a major hurdle for synthetic genetic circuit design. 

%(*) highlight the difference between showing it works on a testbed (earlier sections) and showing it works within a running biological process (later sections) where context and growth rate dynamically change. Growth rate plays against you, chromatin plays against you. These are next important challenges

%(*) more broadly: precision control still difficult – setting the reference point. Models still inappropriate for quantitative design. ML versus physics-based - check Jichi/EDS

%(*) controller parameters uncertain themselves

%(*) sensing and feedback from the state of the cell inside the cell

 \section{ACKNOWLEDGMENT}

  This work was supported in part by  the AFOSR MURI Award  FA9550-22-1-0316.

%\section{Author Information}

%BLA BLA

\bibliographystyle{ieeetr}
\bibliography{Refs}

\end{document}